\documentclass[prx,aps,twocolumn,superscriptaddress,notitlepage]{revtex4-1}

\usepackage [english]{babel}
\usepackage [utf8]{inputenc}

\usepackage{dsfont}
\usepackage{blindtext}
\usepackage{graphicx}
\usepackage{amsmath}
\usepackage{amssymb}
\usepackage{times}
\usepackage{color}
\usepackage{bbm}
\usepackage{bm}
\usepackage{xcolor}
\usepackage{xr}
\usepackage{xfrac}
\DeclareMathAlphabet{\pazocal}{OMS}{zplm}{m}{n}
\usepackage{ulem}
\usepackage{natbib}
\usepackage{subfigure}
\usepackage[colorinlistoftodos, shadow,textsize= footnotesize]{todonotes}
\usepackage[linktocpage=true,colorlinks=true]{hyperref}
\hypersetup{citecolor=magenta,linkcolor=blue,urlcolor=blue}

\newcommand{\mean}[1]{\ensuremath{\langle #1 \rangle}}

\newcommand{\be}{\begin{equation}}
\newcommand{\ee}{\end{equation}}
\newcommand{\ben}{\begin{equation*}}
\newcommand{\een}{\end{equation*}}
\newcommand{\ud}{\mathrm{d}}
\newcommand{\beq}{\begin{eqnarray}}
\newcommand{\eeq}{\end{eqnarray}}

\newcommand{\ii}{{\rm i}}
\newcommand{\cpc}{\cp_{\rm c}}
\newcommand{\evap}{\cp^*}
\newcommand{\lambdadir}{\lambda_z}

\newcommand{\Tmax}{T_{\rm max}}
\newcommand{\Tcross}{T_{\rm cross}}
\newcommand{\DeltaE}{\Delta}		
\newcommand{\neper}{{\rm{e}}}

\newcommand{\sites}{{N}}
\newcommand{\Eunit}{\mathcal{J}}

\newcommand{\bra}[1]{\ensuremath{\langle#1|}}
\newcommand{\ket}[1]{\ensuremath{|#1\rangle}}
\newcommand{\Eins}{\ensuremath{\mathbbm 1}}

\newcommand{\tr}{{\rm Tr}}

\newcommand{\cp}{\lambda}

\begin{document}

\title{Multipartite Entanglement at Finite Temperature}

\author{Marco Gabbrielli}
\affiliation{QSTAR, INO-CNR and LENS, Largo Enrico Fermi 2, I-50125 Firenze, Italy}
\author{Augusto Smerzi}
\affiliation{QSTAR, INO-CNR and LENS, Largo Enrico Fermi 2, I-50125 Firenze, Italy}
\author{Luca Pezz\`e}
\affiliation{QSTAR, INO-CNR and LENS, Largo Enrico Fermi 2, I-50125 Firenze, Italy}

\date{\today}

\begin{abstract}

The interplay of quantum and thermal fluctuations in the vicinity of a quantum critical point characterizes the physics of strongly correlated systems. 
Here we investigate this interplay from a quantum information perspective presenting the universal phase diagram of the quantum Fisher information 
at a quantum phase transition. 
Different regions in the diagram are identified by characteristic scaling laws of the quantum Fisher information with respect to temperature. 
This feature has immediate consequences on the thermal robustness of quantum coherence and multipartite entanglement. 
We support the theoretical predictions with the analysis of paradigmatic spin systems showing symmetry-breaking quantum phase transitions and 
free-fermion models characterized by topological phases.
In particular we show that topological systems are characterized by the survival of large multipartite entanglement, reaching the Heisenberg limit at finite temperature.
 
\end{abstract}

\maketitle
\date{\today}

\section{Introduction} 
\label{SectionIntro}

A quantum information approach to the study of quantum phase transitions (QPTs)~\cite{AmicoRMP2008, EisertRMP2010, ZengARXIV} 
sheds new light on these many-body phenomena~\cite{Sachdev2011} and pushes our understanding 
of the puzzling behavior of strongly-correlated systems~\cite{GhoshNATURE2003, KoppNATPHYS2005, ColemanNATURE2005}
beyond standard methods in statistical mechanics \cite{Huang1987}.  
Entanglement in the ground state of a many-body Hamiltonian $\hat{H}(\lambda) = \hat{H}_0 + \lambda \hat{H}_1$ -- where $\hat{H}_0$ and $\hat{H}_1$ 
are non-commuting operators and $\lambda$ is a control parameter -- has been extensively investigated 
close to a quantum critical point $\cpc$~\cite{AmicoRMP2008, EisertRMP2010, ZengARXIV, OsbornePRA2002, OsterlohNATURE2002, VidalPRL2003, VerstraetePRL2004, WuPRL2004}.
Yet, less is known about the survival of entanglement at finite temperature~\cite{OsbornePRA2002, AmicoEPL2007}, 
including the peculiar quantum critical region that fans out from $\cpc$ 
~\cite{Sachdev2011, ChakravartyPRB1989, SachdevPHYSTODAY2011}. 
This regime is particularly interesting due to the competition of 
thermal and quantum fluctuations~\cite{Sachdev2011, ChakravartyPRB1989, SachdevPHYSTODAY2011} and 
plays a key role in interpreting a wide variety of 
experiments in synthetic matter~\cite{SchroederNAT2000, GrigeraSCIENCE2001, LakeNATMAT2005, LorenzPRL2008, 
DaouNATPHYS2009, ColdeaSCIENCE2010, KinrossPRX2014, KeimerPRB1992, KobayashiPRB1999, CooperSCIENCE2009}.

\begin{figure}[b!]
\centering
\includegraphics[width=0.47\textwidth]{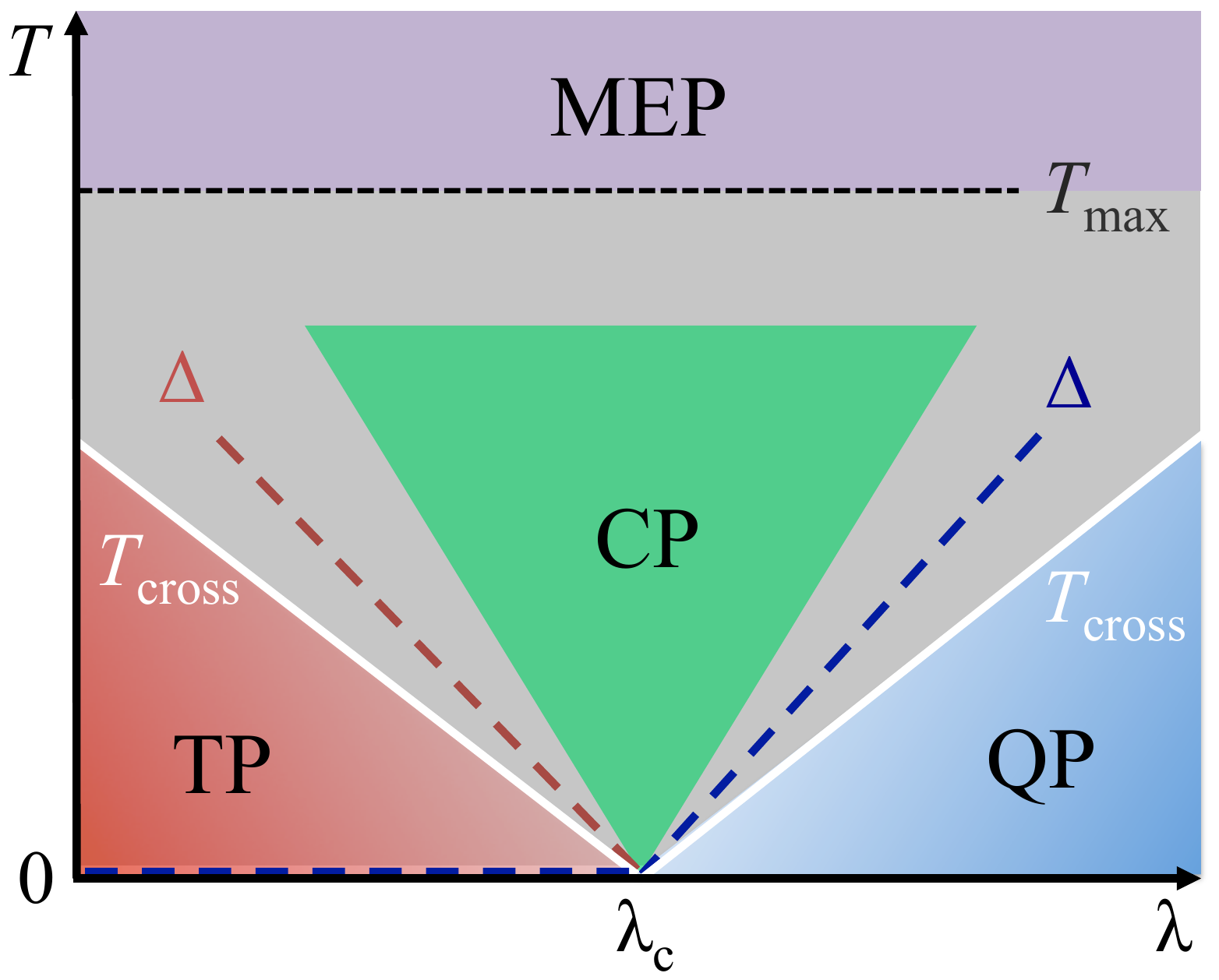}
\caption{
{\bf Schematic general behavior of the scaling of the QFI in the vicinity of a critical point.}
Control parameter $\lambda$ versus temperature $T$ for the QFI of a critical many-body system.
We distinguish four regions depending on the scaling exponent $\beta = d\log F_Q/d \log T$ of the QFI with respect to temperature: 
a quantum plateau (QP), a thermal plateau (TP), a critical plateau (CP) and a maximum entropy plateau (MEP).
QP and TP are defined from the lower bound Eq.~(\ref{QFIfactorization}), showing that the  
QFI remains at least constant ($\beta \geq 0$) up to a crossover temperature $\Tcross$ (white solid line)
of the order of the first nonvanishing gap $\Delta$ in the energy spectrum (dashed line).
The characteristic feature of the TP region is the degeneracy of the ground state:
in the thermodynamic limit, the QFI suddenly decreases from its value at $T=0$ to the plateau value.
In the CP, the QFI follows a scaling law controlled by critical exponents of the model, $\beta = -\Delta_Q/z$, according to Eq.~(\ref{QC}).
For temperatures larger than $\Tmax$ (dotted line) -- approximatively equal to the maximum energy of the spectrum --
the QFI enters the MEP where $\beta=-2$.
In the crossover grey regions the thermal decay is non-universal.} 
\label{fig1}
\end{figure}

Current studies on entanglement in strongly-correlated systems~\cite{AmicoRMP2008, ZengARXIV, EisertRMP2010} 
have mainly focused on bipartite and pairwise entanglement~\cite{HorodeckiRMP2009}. 
This is, however, clearly unsuited to capture the richness of multiparticle correlations
and hardly accessible experimentally in systems of a large number of particles~\cite{IslamNATURE2015}
that are the natural targets of quantum simulators~\cite{ciracNATPHYS2012, GeorgescuRMP2014}. 
Much less attention has been devoted to witnessing multipartite 
entanglement~\cite{GuhneNJP2005, WilmsJSM2012, NakataPRA2009, SadiekJPB2013, MateraPRA2008, HofmannPRB2014, HaukeNATPHYS2016} 
and this has been mainly limited to spin models.
While only few witnesses are known in the literature~\cite{GuhnePR2009},
multipartite entanglement up to hundreds/thousands of spins has been successfully 
detected experimentally in atomic ensembles~\cite{PezzeRMP}.
Among these witnesses, the quantum Fisher information (QFI) has proved to be especially 
suitable~\cite{PezzeRMP, PezzePRL2009, HyllusPRA2012, TothPRA2012, GessnerPRA2016} and it is 
currently attracting considerable interest~\cite{Strobel2014, HaukeNATPHYS2016, PezzePNAS2016, 
RajabpourPRD2017, LiuJPA2013, HaukeNATPHYS2016, PezzePRL2017, ZhangPRL2018, PappalardiJSM2017, GabbrielliARXIV2018}.
The QFI has an appealing operational meaning in terms of statistical speed
of quantum states under external parametric transformations~\cite{PezzePNAS2016, Strobel2014}, 
it extends the class of states detectable by popular methods such as the spin squeezing~\cite{sorensen2001, Strobel2014, LueckeSCIENCE2011, BohnetSCIENCE2016, PezzePRL2009}, 
and it can witness entanglement in spin systems~\cite{MaPRA2009, LiuJPA2013, HaukeNATPHYS2016} 
as well as in free-fermion topological models~\cite{PezzePRL2017, ZhangPRL2018}.
Furthermore, the QFI can be extracted experimentally 
using a statistical distance method~\cite{PezzePNAS2016, Strobel2014}, 
or by a weighted integral of  the dynamic susceptibility across the full spectrum~\cite{HaukeNATPHYS2016}.
Measurable lower bound to the QFI has been extacted experimentally~\cite{Strobel2014, LueckeSCIENCE2011, BohnetSCIENCE2016}
see also Refs.~\cite{FrerotPRB2016, ApellanizPRA2017,MacriPRA2016} for further lower bounds and strategies to extract the QFI. 
The QFI $F_Q[\hat{\rho}, \hat{O}]$ plays a central role in the theory of 
quantum coherence~\cite{StreltsovRMP2017, MarvianPRA2016, MarvianPRA2016b, GirolamiENTROPY2017, LuoPRA2017}: 
it quantifies the coherent extent of a generic state $\hat{\rho}$ over the eigenstates of the operator $\hat{O}$, vanishing if and only if $[\hat{\rho}, \hat{O}]=0$.
Multipartite entanglement is witnessed when the QFI overcomes certain 
finite bounds: as discussed below~\cite{HyllusPRA2012, TothPRA2012}, $F_Q[\hat{\rho}, \hat{O}]>\kappa N$ is 
only achievable if $\hat{\rho}$ contains $(\kappa+1)$-partite entanglement among $N$ parties and $\hat{O}$ is a local operator.

In this manuscript we show that the QFI of a many-body system at thermal equilibrium in the vicinity of a quantum critical point $\cpc$
has the universal behavior shown in Fig.~\ref{fig1}.
At low temperature, the QFI satisfies the inequality
\be \label{QFIfactorization}
\frac{F_Q [\hat{\rho}_T, \hat{O} ]}{ 
F_Q [\hat{\rho}_0, \hat{O} ]} \geq  \tanh^2 \left(\frac{\DeltaE}{2T}\right) 
\frac{\mu(1+\neper^{-\DeltaE/T})}{\mu+\nu \, \neper^{-\DeltaE/T}}.
\ee
Here, $\hat{\rho}_T$ is the thermal state at temperature $T$ (here and in the following the Boltzmann constant is set to 1), 
$\mu$ and $\nu$ indicate the degeneracy of the ground state of energy $E_{\rm gs}$ 
and first excited state of energy $E_{\rm ex}$, respectively, and
$\DeltaE= E_{\rm ex}-E_{\rm gs}$ is the first energy gap in the many-body spectrum. 
Equation (\ref{QFIfactorization}) is valid for $T \lesssim \Delta$ and shows that,
regardless on the microscopical details of the system, the lower bound to $F_Q[\hat{\rho}_T, \hat{O}]$ factorizes 
in a thermal and a quantum contribution.
The thermal decaying function on the right side of the inequality (\ref{QFIfactorization}) only depends on the structure of the 
low-energy spectrum, {\it i.e.}~the energy gap and the degeneracy of the energy eigenstates.
The bound is tight for $T \to 0$, where $F_Q[\hat{\rho}_0,  \hat{O}]$ is the zero-temperature limit of the 
QFI and depends whether the ground state is degenerate or not.

If the ground state is nondegenerate ($\mu=1$), given by the pure state $\ket{\psi_0}$, 
a Taylor expansion of the right-hand side of  
Eq.~(\ref{QFIfactorization}) gives
\be \label{QFIgs1}
\frac{F_Q [\hat{\rho}_T, \hat{O} ]}{F_Q [\ket{\psi_0}, \hat{O} ]} \geq 1 - (3+\nu) \neper^{-\Delta/T} 
+ \mathcal{O}(\neper^{-\Delta/T})^2,
\ee
and shows that the QFI is bounded from below by a constant for $T \lesssim \Tcross$, where $\Tcross \approx \Delta / \log(3+\nu)$.
This defines a {\it quantum plateau} (QP) where the zero-temperature QFI, $F_Q [\ket{\psi_0},  \hat{O}]$,
is insensitive to thermal fluctuations, being protected by the finite energy gap $\Delta$.
In particular, if $\ket{\psi_0}$ hosts multipartite entanglement witnessed by the QFI, such 
multipartite entanglement is robust against temperature for $T \lesssim \Tcross$.
In the following we provide examples of systems characterized by large multipartite entanglement in the ground state 
(even approaching the Heisenberg scaling at finite $T$, see Sec.~\ref{SectionTopological}) 
that is insensitive to small temperatures.

Whenever the ground state is degenerate ($\mu>1$), in the limit $T\to0$ the QFI is given by $F_Q[\hat{\rho}_0, \hat{O}]$, 
where $\hat{\rho}_0$ is the incoherent mixture of the $\mu$ degenerate ground states, see Sec.~\ref{R&M}.
According to Eq.~(\ref{QFIfactorization}), 
\be \label{QFIgs2}
\frac{F_Q [\hat{\rho}_T, \hat{O} ]}{F_Q [\hat{\rho}_0, \hat{O}]} \geq 1 - \bigg(3+\frac{\nu}{\mu}\bigg)  \neper^{-\Delta/T} 
+ \mathcal{O}(\neper^{-\Delta/T})^2.
\ee
Also in this case, the lower bound remains constant for $T \lesssim \Tcross$, where $\Tcross \approx \Delta / \log(3+\nu/\mu)$.
If the ground state becomes degenerate only in the thermodynamic limit, 
this constant value defines a {\it thermal plateau} (TP) where thermal fluctuations strongly affect the 
QFI of the (pure) ground state $\ket{\psi_0}$ outside the thermodynamic limit, but not the QFI of the incoherent mixture $\hat{\rho}_0$.
In other words, the QFI of the ground state $F_Q[\ket{\psi_0},  \hat{O}]$ may be very high -- 
$\ket{\psi_0}$ being given for instance by a maximally entangled state -- but it exponentially 
decays with temperature to a much smaller value $F_Q [\hat{\rho}_0, \hat{O}]$ that remains constant up to $\Tcross$.
In Fig.~\ref{fig1} we schematically plot the case of a typical symmetry-breaking model,
where the TP (matching the ordered phase) and the QP (matching the disordered phase) are found on different sides of the critical point.
Examples of symmetry-breaking models will be discussed in more details in Sec.~\ref{SectionSBQPT}.
In the absence of ground-state degeneracy, the TP is absent and the QP is found on both sides of the critical point.
This behavior is found for topological closed chains, as shown in Sec.~\ref{SectionTopological}. 

At finite temperature and for values of $\lambda$ around the critical point $\cpc$, a scaling hypothesis for the dynamical susceptibility~\cite{HaukeNATPHYS2016}
predicts 
\be \label{QC}
\frac{F_Q[\hat{\rho}_T, \hat{O}]}{N} \sim T^{-\Delta_Q/z}.
\ee 
Here, $N$ is the total number of parties in the system ({\it e.g.} the total number of spins), 
$\Delta_Q$ is the exponent that characterizes the finite-size scaling of the QFI 
with respect to $N$ at $T=0$ and $\lambda=\cpc$, 
{\it i.e.}~$F_Q[\ket{\psi_0},  \hat{O}]/N \sim N^{\Delta_Q/d}$~\cite{HaukeNATPHYS2016}, 
and $z$ is the dynamical critical exponent.
We thus identify a region of parameters in the vicinity of the critical point ($T>0$) 
that we call {\it critical plateau} (CP)
where the QFI follows the scaling behavior Eq.~(\ref{QC}) as a function of temperature.
In general, we expect that the CP extends for $T \gg \vert \lambda - \cpc \vert^{\nu z}$, where $\nu$ is the correlation-length critical exponent.
This region matches a quantum critical regime~\cite{Sachdev2011, SachdevPHYSTODAY2011} where 
the scaling behavior of a quantum coherence measure, 
the QFI, at finite temperature is controlled by critical exponents of the transition.
In Fig.~\ref{fig1} the CP is schematically represented as a triangular region.
The CP is separated from the TP and QP by a model-dependent smooth decay for $T \approx \Tcross$.

Finally, for temperatures of the order of the interaction energy scale of the system, 
no multipartite entanglement is witnessed by the QFI. 
Moreover, for temperatures larger than the maximum energy of the spectrum, the QFI decays as 
\be \label{MEP}
F_Q[\hat{\rho}_T, \hat{O}] \sim T^{-2}.
\ee
This defines a fourth plateau that we identify as 
{\it maximum entropy plateau} (MEP). 
In this regime, all eigenstates are approximatively equally populated. 

It is worth clarifying that the operator $\hat{O}$ in Eqs.~(\ref{QFIfactorization})-(\ref{QFIgs2}) and (\ref{MEP}) is arbitrary, while Eq.~(\ref{QC})
holds for the order parameter of the quantum phase transition.

The manuscript is organized as follows:
in Sec.~\ref{R&M}, we provide a detailed derivation of the equations discussed above.
In the remaining sections, we draw the finite-temperature phase diagram of the QFI in hallmark systems, recovering the schematic behavior shown in Fig.~\ref{fig1}.
In Sec.~\ref{SectionSBQPT} we study symmetry-breaking QPTs, 
focusing on the Ising model and the bosonic Josephson junction, 
while in Sec.~\ref{SectionTopological} we consider topological QPTs, 
in particular the ones in the Kitaev chain with variable-range pairing. 
Finally, discussions and conclusions are reported in Sec.~\ref{outlook}.

\section{Methods and Results}
\label{R&M}

\subsection{Quantum Fisher information, multipartite entanglement and quantum coherence}
\label{kentB}

\begin{figure}[t!]
\centering
\includegraphics[width=0.48\textwidth]{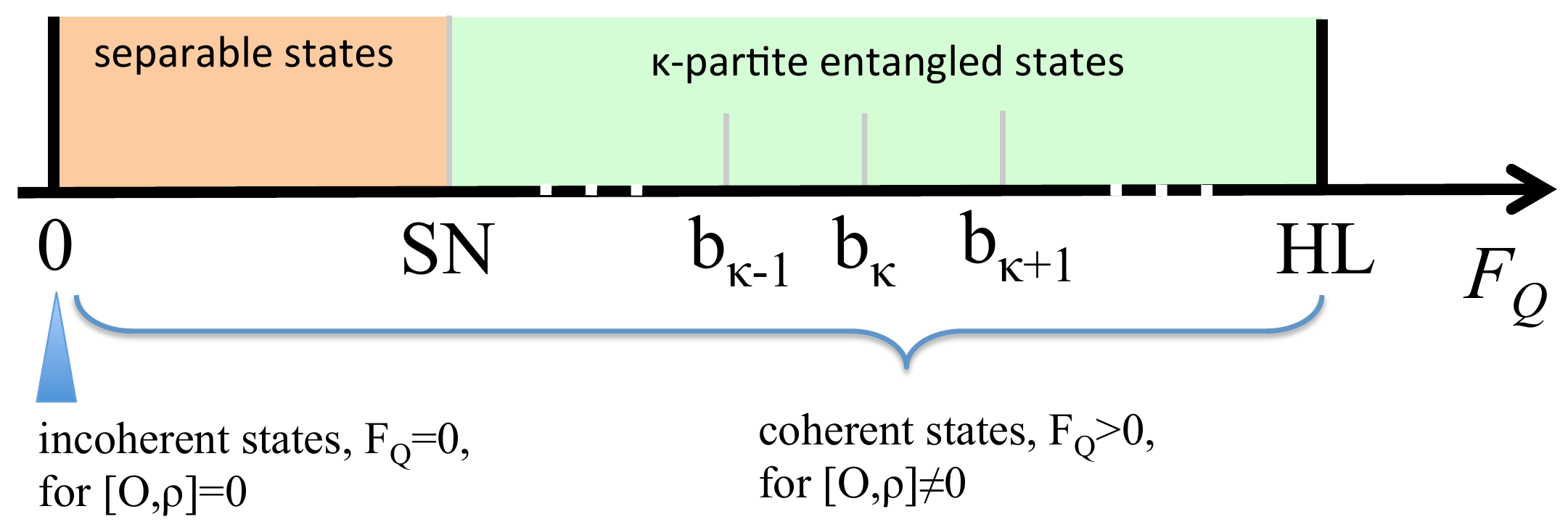}
\caption{
{\bf Bounds of the QFI.}
For unitary phase-encoding transformations, $F_Q[\hat{\rho}, \hat{O}] =0$ if and only if the state is incoherent. 
Among quantum coherent states, $F_Q[\hat{\rho}, \hat{O}] > 0$, we can find bounds to the QFI depending on the entanglement properties of the state: 
$F_Q[\hat{\rho}, \hat{O}] \leq b_1$ for separable states [orange region, where $b_1$ is also indicated as shot-noise (SN) limit], 
$F_Q[\hat{\rho}, \hat{O}] \leq b_\kappa$ for $\kappa$-partite entangled states with $1\leq \kappa\leq N-1$ [green region], and 
$F_Q[\hat{\rho}, \hat{O}] \leq b_N$ for all possible states [where $b_N$ is also indicated as Heisenberg limit (HL)].
The bounds $b_\kappa$ depend, in general, on the operator $\hat{O}$.
} 
\label{fig2}
\end{figure}

The QFI quantifies the distinguishability between nearby quantum states 
$\hat{\rho}$ and $\hat{\rho}_\varphi$ related by an arbitrary transformation depending on the parameter $\varphi$.
The Uhlmann fidelity~\cite{Uhlmann1976} between $\hat{\rho}$ and $\hat{\rho}_\varphi$
is $\mathcal{F}[\hat{\rho}, \hat{\rho}_\varphi]  = \tr[\sqrt{\sqrt{\hat{\rho}} \hat{\rho}_\varphi \sqrt{\hat{\rho}}}] = 1 - 
\tfrac{1}{8} F_Q[\hat{\rho}_\varphi] \varphi^2 + \mathcal{O}(\varphi^3)$, where 
$F_Q[\hat{\rho}_\varphi]$ is the QFI.
In terms of the spectral decomposition $\hat{\rho}_\varphi = \sum_k p_k \ket{k} \bra{k}$ (with $p_k \geq 0$ and $\sum_k p_k =1$)
we have $F_Q[\hat{\rho}_\varphi]  = \sum_{k,k'} \tfrac{2}{p_k+p_{k'}} \vert \bra{k} \partial_\varphi \hat{\rho}_\varphi \ket{k'} \vert^2$ provided that $p_k+p_{k'}\neq0$.
The QFI has key mathematical properties \cite{BraunsteinPRL1994, petz2010, pezze2014, toth2014}, that allow the derivation of 
relevant bounds, see Fig.~\ref{fig2}:

\begin{description}
\item[i] {\it Convexity.} The QFI is nonnegative and convex in the state:
\be \label{Fqkent}
F_Q\Big[\sum_{i} q_i \hat{\rho}_{\varphi}^{(i)} \Big] \leq \sum_i q_i  F_Q[\hat{\rho}_{\varphi}^{(i)}],
\ee
for any state $\hat{\rho}_{\varphi}^{(i)}$ and $q_i \geq 0$.
\item[ii] {\it Additivity.} The QFI is additive under tensor product:
\be
F_Q[\hat{\rho}_{\varphi}^{(1)} \otimes \hat{\rho}_{\varphi}^{(2)}] = 
F_Q[\hat{\rho}_{\varphi}^{(1)}] + F_Q[\hat{\rho}_{\varphi}^{(2)}]. 
\ee
\item[iii] {\it Monotonicity.} The QFI always decreases under arbitrary 
parameter-independent completely positive trace-preserving map $\Lambda$:
\be \label{monoton}
F_Q[\Lambda(\hat{\rho}_\varphi)] \leq F_Q[\hat{\rho}_\varphi], 
\ee
with equality for $\varphi$-independent unitary transformations. 
\end{description}
In the following we will restrict to unitary transformations, 
$\hat{\rho}_\varphi = \neper^{-\ii\varphi\hat{O}} \, \hat{\rho} \, \neper^{\ii\varphi\hat{O}}$
where $\hat{O}$ is a generic Hermitian operator that we will specify below. 
The unitary transformation only evolves the eigenstates of $\hat{\rho}$ and leave its eigenvalues unchanged. 
For unitary transformations, the QFI has the following further properties:
\begin{description}
\item[iv] The QFI satisfies 
\be
F_Q[\neper^{-\ii\varphi\hat{O}} \, \hat{\rho} \, \neper^{\ii\varphi\hat{O}}] \equiv F_Q[\hat{\rho}, \hat{O}] \leq 4 (\Delta \hat{O})^2_{\hat{\rho}},
\ee 
with equality for pure states. 
\item[v] The QFI vanishes if and only if $\hat{\rho}$ and $\hat{O}$ can be diagonalized simultaneously:
\be 
F_Q[\hat{\rho}, \hat{O}] =0 \ \Leftrightarrow \ [\hat{\rho},\hat{O}]=0. 
\ee 

\end{description}

\subsubsection{QFI and quantum coherence}

The coherence of a quantum state $\hat{\rho}$ is defined from its distinguishability with respect to the set of states that are diagonal 
in a given basis~\cite{StreltsovRMP2017}. 
Here such a basis is given by the eigenstates of the operator $\hat{O}$, and incoherent states are those satisfying $[\hat{\rho}, \hat{O}]=0$.
In addition to the properties ({\bf i}) and ({\bf v}), the QFI does not increase under operations that conserve $\hat{O}$, namely
$F_Q[\Lambda_C[\hat{\rho}], \hat{O}] \leq F_Q[\hat{\rho}, \hat{O}]$ for maps $\Lambda_C$ satisfying
$\Lambda_C[ \neper^{-\ii\varphi\hat{O}} \, \hat{\rho} \, \neper^{\ii\varphi\hat{O}} ]= \neper^{-\ii\varphi\hat{O}}  \Lambda_C[\hat{\rho}] \neper^{\ii\varphi\hat{O}}$.
These properties make the QFI a reliable measure of asymmetry~\cite{MarvianPRA2016} --
a broad notion of quantum coherence~\cite{StreltsovRMP2017, MarvianPRA2016b}.
Physically, the concept of asymmetry quantifies how much a state $\hat{\rho}$ satisfying $[\hat{\rho}, \hat{O}] \neq 0$ 
changes when applying the unitary transformation $\neper^{-\ii\varphi\hat{O}}$. 
The changes in the state can be used to estimate the 
phase $\varphi$ with a nonvanishing sensitivity $(\Delta \varphi)^2 = 1/F_Q[\hat{\rho}, \hat{O}]$~\cite{PezzeRMP, GiovannettiNATPHOT2011} 
in a sensor implementing the transformation $\neper^{-\ii\varphi\hat{O}}$. \\

\subsubsection{QFI and multipartite entanglement}

The key property that makes the QFI a multipartite entanglement witness 
\cite{PezzePRL2009, HyllusPRA2012, TothPRA2012, PezzePNAS2016}
is the convexity in the state [property ({\bf i}) above].
We recall that a pure state is $\kappa$-partite entangled if it can be written as 
$\ket{\psi_{\kappa\textrm{-}{\rm ent}}} = \otimes_{j} \ket{\psi_j}$~\cite{GuhnePR2009}, 
where $\ket{\psi_j}$ is a state of $N_j \leq \kappa$ parties
(with $\sum_j N_j = N$, $N$ being the total number of parties in the system) that does not factorize.
In other words, $\kappa$-partite entanglement indicates the number of parties in the largest nonseparable subset.
$\kappa$-partite entangled states form a convex set and we can indicate with 
$\hat{\rho}_{\kappa\textrm{-}{\rm sep}} = \sum_i p_i \ket{\psi_{\kappa\textrm{-}{\rm ent}}^{(i)}} \bra{\psi_{\kappa\textrm{-}{\rm ent}}^{(i)}}$ 
a generic element of the ensemble.
As a consequence of Eq.~(\ref{Fqkent}), every (pure or mixed) $\kappa$-partite entangled state
satisfies $F_Q[\hat{\rho}_{\kappa\textrm{-}{\rm sep}}] \leq b_{\kappa, \hat{O}}$, where 
\be \label{bk}
b_{\kappa, \hat{O}}= 4 \max_{\ket{\psi_{\kappa\textrm{-}{\rm ent}}}} (\Delta O)^2_{\ket{\psi_{\kappa\textrm{-}{\rm ent}}}}.
\ee
The maximization is done over all possible $\kappa$-separable pure states 
and we have used $F_Q\big[\ket{\psi}, \hat{O} \big] = 4 (\Delta \hat{O})^2_{\ket{\psi}}$.
A theoretical challenge is to calculate the multipartite entanglement bounds (\ref{bk}) for a 
given operator $\hat{O}$, which might be local \cite{PezzePRL2009, HyllusPRA2012, TothPRA2012} or nonlocal \cite{PezzePNAS2016}. 
The choice of the operator involved in the calculation of the QFI leads to different entanglement bounds $b_{\kappa, \hat{O}}$.
While there is no known systematic method to choose the optimal operator $\hat{O}$ 
({\it i.e.}~the one that allows the detection of the largest class of states),
an ``educated guess'' based on some knowledge of the system 
allows the corresponding QFI to witness multipartite entanglement close to QPTs for different models.
For instance, in models showing symmetry-breaking QPTs, the transition is characterized by the divergence of fluctuations of a local order parameter.
We thus expect a large QFI at criticality when $\hat{O}$ is given by the order parameter of the transition \cite{HaukeNATPHYS2016}.
In spin models such as the Ising and the bosonic Josephson junction models 
this is a collective spin operators (given by the sum of Pauli matrices).
In this case we have $b_{\kappa, \hat{O}} = s \kappa^2 + r^2 \approx N \kappa$ \cite{HyllusPRA2012, TothPRA2012}, 
where $s = \lfloor N/\kappa \rfloor$ is the largest integer smaller or equal than $N/\kappa$ and $r = N - s\kappa$.
A QFI larger than this bound witnesses $(\kappa+1)$-partite entanglement between spin-1/2 particles.
On the contrary, topological QPTs are not detected by a local order parameter. 
In order to witness multipartite entanglement in topological models 
it is thus necessary to calculate the QFI with respect to nonlocal operators.
For the one-dimensional short-range Kitaev chain discussed below, an optimal choice of operator is 
suggested by the correspondence, via the Jordan-Wigner transformation, to the Ising model.  
Indeed, the QFI is able to detect multipartite entanglement in  a topological system~\cite{PezzePRL2017} 
when choosing, as operator $\hat{O}$,
the Jordan-Wigner transformation of the local order parameter for the Ising chain (see below). 
Furthermore, this choice leads to the same multipartite entanglement bounds $b_{\kappa,\hat{O}} = s \kappa^2 + r^2 \approx N \kappa$ \cite{PezzePRL2017}.

\subsection{Quantum Fisher information of thermal states}
\label{QFIth}

We consider a generic thermal state at canonical equilibrium, $\hat{\rho}_T = \neper^{-\hat{H}/T} / \mathcal{Z}$, 
where $\hat{H}$ is the many-body Hamiltonian with eigenenergies $E_n$ and corresponding eigenstates $\ket{\psi_n}$, 
$T$ is the temperature, $p_n = \neper^{-E_n/T} / \mathcal{Z}$ 
and $\mathcal{Z} = \sum_n \neper^{-E_n/T}$ is the partition function.
The QFI of $\hat{\rho}_T$, calculated with respect to the operator $\hat{O}$, is 
\be \label{QFImixed1}
F_Q[\hat{\rho}_T, \hat{O}] = 2 \sum_{n,m} 
\frac{(p_n - p_m)^2}{p_n + p_m} 
\,|\hat{O}_{n,m}|^2,
\ee
where $\hat{O}_{n,m} = \langle \psi_n |\hat{O}| \psi_m \rangle$.
Notice that $F_Q[\hat{\rho}_T, \hat{O}] \leq 4 (\Delta \hat{O})^2$ at all temperatures. 
Equation (\ref{QFImixed1}) can be rewritten as 
\be \label{QFImixed2}
F_Q[\hat{\rho}_T, \hat{O}] = 4 \sum_{n} p_n \big( \Delta \hat{O} \big)^2_{\ket{\psi_n}} 
- 8 \sum_{\substack{n,m \\ n\neq m}}
\frac{ p_n p_m}{p_n + p_m} \,|\hat{O}_{n,m}|^2. 
\ee
Computing the QFI using Eqs.~(\ref{QFImixed1}) or (\ref{QFImixed2}) requires the diagonalization of the full Hamiltonian $\hat{H}$.
A calculation using a limited manifold of eigenstates ({\it i.e.}~in a Hilbert subspace given by the most populated states at temperature $T$)
only leads to approximate results, since the matrix element $\hat{O}_{n,m}$ may couple to energy eigenstates outside the manifold.
The calculation of the QFI in a Hilbert subspace (as discussed below for the two-mode approximation) leads to 
accurate results provided that coupling terms between the subspace and the rest of the Hilbert space induced by the operators $\hat{O}$ are negligible. 

The QFI (\ref{QFImixed1}) can also be rewritten in the useful form~\cite{HaukeNATPHYS2016} 
\be \label{FQS0}
F_Q[\hat{\rho}_T, \hat{O}] = \frac{4 \hbar }{\pi} \int_{0}^{+\infty} \ud{\omega} \tanh \left(\frac{\hbar \omega}{2T}\right) {\rm Im}\chi_O(\omega,T),
\ee
where ${\rm Im} \chi_O(\omega,T) = \pi \sum_{n,m} (p_m - p_n) |\hat{O}_{n,m}|^2 \delta(\hbar \omega - \hbar \omega_{n,m})$
is the imaginary part of the dynamical susceptibility $\chi_O$, 
and $\hbar \omega_{n,m}  = E_n - E_m$.
Using the fluctuation-dissipation relation ${\rm Im} \chi_O(\omega,T) = \frac{1}{\hbar}\tanh\big(\tfrac{\hbar \omega}{2T}\big) S_O(\omega,T)$
we can write 
\be \label{FQS1}
F_Q[\hat{\rho}_T, \hat{O}] = \frac{2}{\pi} \int_{-\infty}^{+\infty} \ud{\omega} \tanh^2\left(\frac{\hbar \omega}{2T}\right)\mathcal{S}_O(\omega,T),
\ee
where 
$\mathcal{S}_O(\omega,T) = \int_{-\infty}^{+\infty}\ud{t}\,\neper^{\ii\omega{t}}\, {\rm Re} \langle \hat{O}(t) \hat{O}\rangle  
=  \pi \sum_{n,m} (p_m + p_n) |\hat{O}_{n,m}|^2 \delta(\omega-\omega_{n,m})$
is the dynamic structure factor,  
$\hat{O}(t)=\neper^{\ii\hat{H}t/\hbar}\hat{O}\neper^{-\ii\hat{H}t/\hbar}$, and we have used the property 
$\mathcal{S}_O(-\omega,T) = \mathcal{S}_O(\omega,T)$. 
Equation~(\ref{FQS1}) can thus be rewritten as
\be \label{FQS2}
F_Q[\hat{\rho}_T, \hat{O}] = 4 \langle \hat{O}^2 \rangle - 8 (T/\hbar)^2
\int_{-\infty}^{+\infty} \ud{t} \, \frac{{\rm Re} \langle\hat{O}(t)\hat{O}\rangle }{\sinh (\pi T t /\hbar )/t},
\ee
which shows that the QFI can be calculated 
from the knowledge of the time correlation functions $\langle\hat{O}(t)\hat{O}\rangle $.
These are known, for instance, in the Ising model for certain operators~\cite{DerzhkoPRB1997} without requiring the full 
diagonalization of the Hamiltonian \cite{NOTA01}.

\subsubsection{Zero-temperature case}

At zero temperature, the QFI becomes  
\be \label{QFI0degerate}
F_Q[\hat{\rho}_0, \hat{O}] = \frac{4}{\mu} \Bigg( \sum_{d=1}^{\mu} \big( \Delta \hat{O} \big)^2_{\ket{\psi_0^{(d)}}} - \sum_{\substack{ d,d'=1 \\ d \neq d'}}^{\mu}\big|\bra{\psi_0^{(d)}}\hat{O}\ket{\psi_0^{(d')}}\big|^2 \Bigg),
\ee
in case the ground state has a degeneracy $\mu$ 
(we have indicated as $\ket{\psi_0^{(d)}}$ the degenerate eigenstates, with $d=1,...\,\mu$), 
such that $\hat{\rho}_0 = \frac{1}{\mu}\sum_{d=1}^{\mu}\ket{\psi_0^{(d)}}\bra{\psi_0^{(d)}}$, 
and reduces to $F_Q[\ket{\psi_0}, \hat{O}] = 4 \big( \Delta \hat{O} \big)^2_{\ket{\psi_0}}$ in absence of degeneracy ($\mu=1$).

\subsubsection{Low-temperature limit and two-mode approximation}

Here we demonstrate the inequality~(\ref{QFIfactorization}).
Let us consider, for simplicity, a nondegenerate spectrum: the equations that we will obtain in this 
section can be straightforwardly extended to the degenerate case. 
At low temperature $T \lesssim \DeltaE$, we can neglect the population of high-energy eigenstates 
({\it i.e.}~taking $p_n=0$ for $n \geq 2$).
In this case, using the completeness relation $\sum_n \ket{\psi_n} \bra{\psi_n} = \Eins$, Eq.~(\ref{QFImixed1}) becomes
\beq \label{QFItwomode1}
F_Q[\hat{\rho}_T, \hat{O}] & = & \frac{(p_1 - p_0)^2}{p_1 + p_0} F_Q[\hat{\rho}_0, \hat{O}] + 
4 p_1 \sum_{m\neq0,1} \vert \hat{O}_{1,m} \vert^2 + \nonumber \\
& + & 4 \bigg( p_0 - \frac{(p_1 - p_0)^2}{p_1 + p_0} \bigg)
\sum_{m\neq0,1} \vert \hat{O}_{0,m} \vert^2 .
\eeq
Notice that the second and third terms in Eq.~(\ref{QFItwomode1}) are always positive (at all temperatures), 
which implies 
\be 
\frac{F_Q[\hat{\rho}_T, \hat{O}]}{F_Q[\hat{\rho}_0, \hat{O}]} \geq \frac{(p_1 - p_0)^2}{p_1 + p_0} = \tanh^2 \left(\frac{\Delta}{2T}\right).
\ee
We thus obtain the inequality (\ref{QFIfactorization}) from which we derive Eqs.~(\ref{QFIgs1}) and (\ref{QFIgs2}).
The inequality is tight in the limit $T\to 0$, when $p_1 = 0$.
It is also tight at all temperature if and only if $\vert \hat{O}_{0,m} \vert, \vert \hat{O}_{1,m} \vert= 0$ for all $m \geq 2$, 
{\it i.e.}~when the operator $\hat{O}$ only couples the ground and the first excited state.

\subsubsection{Quantum critical scaling}
\label{CriticalScaling}

The scaling behavior in Eq.~(\ref{QC}) follows from a standard scaling hypothesis for the dynamical susceptibility, see for instance~\cite{Mussardo, Tauber2014}:
\be \label{chiansatz}
{\rm Im}\chi_O(\omega,T) = \chi \, \phi_O(\hbar\omega/T, T/\Delta, L/\xi),
\ee
where $\chi$ is the static susceptibility of the operator $\hat{O}$ with respect to a coupled field,
$\phi_O$ is a suitable scaling function, 
$\xi$ is the correlation length and $L = N^{1/d}$ is the linear system size, 
being $d$ the system dimension. 
Inserting Eq.~(\ref{chiansatz}) into Eq.~(\ref{FQS0}) we obtain
\be \label{scal1}
F_Q[\hat{\rho}_T, \hat{O}] = \chi N \Delta \, g(T/\Delta, L/\xi),
\ee
where $g\big(\tfrac{T}{\Delta}, \tfrac{L}{\xi}\big) = \tfrac{4}{\pi} \tfrac{T}{\Delta} \int_{0}^{+\infty} \ud{x} \tanh(\tfrac{x}{2}) \phi_O(x, \tfrac{T}{\Delta}, \tfrac{L}{\xi})$.
We now take into account that, close enough to the critical point, 
$\Delta \sim \delta^{z \nu}$, $\xi \sim \delta^{-\nu}$ and $\chi \sim \delta^{-\gamma}$, 
where $\delta = \vert \lambda - \cpc \vert$ and $z$, $\nu$ and $\gamma$ are critical exponents.
Under coarse-graining transformation on the system, lengths scale as $l \to l'=b^{-1}l$ 
($b>1$), while $N \to N'=b^{-d}N$.
A dimensional analysis reveals that $\Delta \to \Delta'=b^{z}\Delta$ and $\delta \to \delta'=b^{1/\nu}\delta$,
whereas the scaling function $g$ only depends on adimensional variables and does not scale under length rescaling.
Thus, under coarse graining the QFI transforms according to 
\be \label{scal2}
F_Q[\hat{\rho}_T, \hat{O}]/N = b^{\gamma/\nu-z} \, h(b^{1/\nu}\delta, b^{z}T, b^{-1}L),
\ee
with $h$ a suitable scaling function.
The behavior of the QFI with respect to relevant quantities can be extracted by setting
the dominant rescaling factor $b$ up to which the scale invariance of the system is preserved.

At small temperatures $T\ll\Delta$, no significant length scale is induced by temperature. 
Sufficiently far from criticality, $\xi \ll L$ and scale invariance is preserved up to $b\sim\xi$. 
Equation~(\ref{scal2}) then implies $F_Q[\hat{\rho}_T, \hat{O}] \sim \delta^{z\nu-\gamma}$ for $T \ll \delta^{z\nu}$~\cite{HaukeNATPHYS2016}.
Conversely, at the critical point $\delta \ll L^{-1/\nu}$, the constituents of the system are correlated on a scale $\xi \gg L$: 
the system experiences finite-size effects and it remains scale invariant up to $b \sim L$.
Equation~(\ref{scal2}) gives the scaling of the QFI with $N$ for $T \ll L^{-z}$~\cite{HaukeNATPHYS2016}:
$F_Q[\hat{\rho}_T, \hat{O}]/N \sim N^{\Delta_Q/d}$, 
where we have used the Fisher relation $\gamma/\nu = 2 -\eta$ and defined $\Delta_Q = 2-\eta-z$.

On the contrary, thermal fluctuations dictate a dominant length scale if $T^{1/z} \gg L^{-1}, \xi^{-1}$.
Scale invariance is expected to be broken at the scale $b \sim T^{-1/z}$. 
Thus, Eq.~(\ref{scal2}) provides the scaling of the QFI with temperature valid for $T\gg\Delta$~\cite{HaukeNATPHYS2016}:
$F_Q[\hat{\rho}_T, \hat{O}] \sim T^{-\Delta_Q/z}$, namely Eq.~(\ref{QC}).
 
\subsubsection{High-temperature limit}

For very large temperature, $T \gtrsim \Tmax = \max_n E_n$, 
we can expand $\neper^{-E_n/T} \approx 1 - E_n/T + \mathcal{O}(E_n/T)^2$. 
Equation~(\ref{QFImixed1}) becomes 
\be
F_Q[\hat{\rho}_T, \hat{O}]  \propto \frac{1}{T^2} \sum_{n,m} (E_n - E_m)^2 \vert \hat{O}_{n,m} \vert^2,
\ee 
which predicts a universal $1/T^2$ scaling. 
In the limit $T \to \infty$ we have $\hat{\rho}_T \propto \Eins$: it commutes with $\hat{O}$ and 
we find $F_Q[\hat{\rho}_T,\hat{O}] \allowbreak = \allowbreak 0$.

\begin{figure*}[t!]
\includegraphics[width=\textwidth]{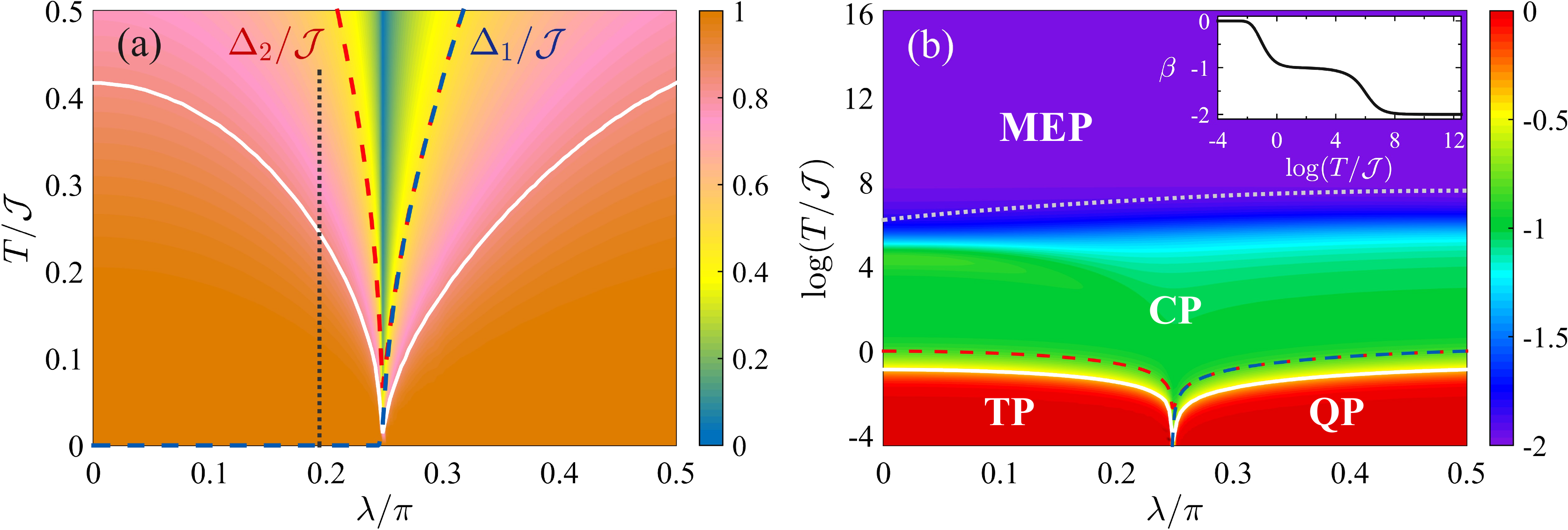} 
\caption{
{\bf Phase diagram of the BJJ model.}
(a) QFI normalized to its low-temperature value, $F_Q[\hat{\rho}_T]/F_Q[\hat{\rho}_0]$ (color scale) 
as a function of $\lambda$ and $T$.
The region where the low-temperature behavior survives is highlighted by the orange color. 
The black dotted line at $\lambda\approx\mathrm{atan}(1/\sqrt{2})$ separates the regions where the optimal parameter is $\hat{S}_x$ (on the left) and $\hat{S}_z$ (on the right).
(b) Scaling coefficient $\beta = d \log F_Q[\hat{\rho}_T] / d \log T$ (color scale)
as a function of $\lambda$ and $T$.
The dotted line is the upper bound of the spectrum, $\max_n E_n$.
The inset shows $\beta$ as a function of $T$ at $\lambda/\pi=0.4$: the different plateaus are clearly visible.
In both panels $N=2000$, 
the solid white curve is the crossover temperature $\Tcross(\cp)$ 
following the energy gaps $\DeltaE_1$ (dashed blue line) and $\DeltaE_2$ (dashed red line).
} 
\label{fig3}
\end{figure*}

\subsubsection{Thermalization in a subspace of the full Hamiltonian}

All the above equations and the analysis in the following sections implicitly assumes 
thermal equilibrium in the full Hilbert space. 
However, it might be possible to have a thermalization only in a Hilbert subspace generated, 
for instance, by a finite subset of the eigenstates of the full Hamiltonian.
This scenario may arise from a metastable equilibrium due to different thermalization time scales of different Hilbert subspaces. 
In this case, $\hat{\rho}_T = \sum_{n} q_n \ket{\psi_n} \bra{\psi_n}$, 
where $q_n \neq 0$ if $n \in \mathcal{H}'$ and $q_n=0$ otherwise, where 
$\mathcal{H}'$ is a subspace of the full Hilbert space $\mathcal{H}$ with a basis given by the states $\ket{\psi_n}$. 
In this case, the QFI writes 
\be \label{Fsub}
F_Q[\hat{\rho}_T, \hat{O}] = 4 \! \sum_{n \in \mathcal{H}'} 
q_n (\Delta \hat{O})^2_{\ket{ \psi_n }} \, - \, 8 \!\!\!\! \sum_{\substack{n,m \in \mathcal{H}' \\ n\neq m}}
\frac{q_n q_m}{q_n + q_m} \vert \hat{O}_{n,m}\vert^2.
\ee
Interestingly, the second term in Eq.~(\ref{Fsub}) can vanishes. 
This occurs when $\vert \hat{O}_{n,m} \vert =0$ $\forall n,m \in \mathcal{H}'$,
{\it i.e.} when the operator $\hat{O}$ does not couple states within the subspace $\mathcal{H}'$.
Notice that $\hat{O}$ may couple states in $\mathcal{H}'$ with states outside this subspace,
but the latter are not populated before the phase-encoding transformation and do not enter into Eq.~(\ref{Fsub}).
In this case, the QFI reduces to 
\be
F_Q[\hat{\rho}_T, \hat{O}] = 4 \sum_{n \in \mathcal{H}'} 
q_n \langle \psi_n |\hat{O}^2 | \psi_n \rangle.
\ee
This equation predicts results that are completely different from the ones discussed above.
For instance, if the excited states are characterized by values of $\langle \psi_n |\hat{O}^2 | \psi_n \rangle$ that are larger than 
those for the low-lying states, then $F_Q[\hat{\rho}_T, \hat{O}]$ may increase with temperature. 
Furthermore, in the large-$T$ limit, taking $\hat{\rho}_T \propto \Eins$,
we obtain that the QFI saturates a finite constant value, 
$F_Q[\hat{\rho}_T, \hat{O}] \propto \sum_{n \in \mathcal{H}'} \langle \psi_n |\hat{O}^2 | \psi_n \rangle$. 
In other words, we may have, in this case, that multipartite entanglement increases with temperature and 
remains large even at high temperature, as recently noticed in a spin-1 system~\cite{KajtochPRA2018}.

\section{Applications: symmetry breaking QPTs} 
\label{SectionSBQPT}

In the following we witness multipartite entanglement at finite temperature in  
two paradigmatic models exhibiting a symmetry-breaking QPT.
We first discuss the bosonic Josephson junction (BJJ) model, 
as it allows for analytical calculations of the QFI at zero as well as finite temperature for large particle numbers.
We then focus on the Ising model in a transverse field, which is a common testbed of quantum criticality. 
The BJJ model can be used to describe a Bose gas in two hyperfine levels coupled 
by a microwave field~\cite{ZiboldPRL2010, Strobel2014} or in a double-well potential 
in the tunneling regime~\cite{EsteveNATURE2008, BerradaNATCOMM2013, TrenkwalderNATPHYS2016, SpagnolliPRL2017},
whereas the Ising model has been realized experimentally with ultracold atoms in an optical lattice \cite{simon2011}, 
trapped ions~\cite{KimNJP2011, BrittonNATURE2012, JurcevicNATURE2014, ZhangNATURE2017}
and solid-state platforms \cite{ColdeaSCIENCE2010, KinrossPRX2014, halg2015}.

\subsection{BJJ model}
\label{SubSectionBJJ}

The BJJ consists of $N$ interacting bosonic particles 
occupying two weakly-coupled modes $\ket{a}$ and $\ket{b}$~\cite{smerzi1997, milburn1997, PezzeRMP}, 
{\it e.g.} two internal levels of an atom or two wells of an external trapping potential.
The system is described by the Hamiltonian
\be \label{HLMG}
\frac{\hat{H}_{\rm BJJ}}{\Eunit} = -\frac{1}{N} \cos\cp \,\hat{S}_z^2 + \sin\cp\,\hat{S}_x
\ee
where
$\hat{S}_x = (\hat{a}^\dag \hat{b} + \hat{b}^{\dag} \hat{a})/2$, 
$\hat{S}_y = (\hat{a}^\dag \hat{b} - \hat{b}^{\dag} \hat{a})/2\ii$ and 
$\hat{S}_z = (\hat{a}^\dag \hat{a} - \hat{b}^{\dag} \hat{b})/2$ satisfy SU(2) commutation relations, 
$\hat{a}$ and $\hat{b}$ ($\hat{a}^\dag$ and $\hat{b}^\dag$) are bosonic annihilation (creation) 
operators for the $\ket{a}$ and $\ket{b}$ modes, respectively. 
The coefficient $\Eunit$ denotes the characteristic energy scale of the system.
The control parameter $\cp\in[0,\pi/2]$ rules the interplay between 
particle-particle interaction, described by $\hat{S}_z^2$, and the linear coupling term, given by $\hat{S}_x$. 
In the thermodynamic limit $N\to \infty$, and for $T=0$, Eq.~(\ref{HLMG}) exhibits a QPT at $\cpc=\pi/4$
between a quantum paramagnetic phase (for $\cpc<\cp\leq\pi/2$) 
and a ferromagnetic long-range ordered phase (for $0\leq\cp<\cpc$)~\cite{botet1982, botetPRB1983, NOTA02}. 

In the following, we calculate the QFI for a thermal state $\hat{\rho}_T$ and optimize with respect to the operators $\hat{O} = \hat{S}_{x,y,z}$.
The optimization procedure consists in  taking the maximum eigenvalue,
$F_Q[\hat{\rho}_T]=\max\mathrm{eigval}\,\mathbb{F}_Q[\hat{\rho}_T]$, of the matrix
\be \label{QFIoptimization}
\mathbb{F}_Q^{kl}[\hat{\rho}_T] = 2 \sum_{n,m} 
\frac{(p_n-p_m)^2}{p_n+p_m}
\,\bra{\psi_n}\hat{S}_k\ket{\psi_m}\bra{\psi_m}\hat{S}_l\ket{\psi_n}
\ee
with $k,l=x,y,z$~\cite{HyllusPRA2010}.
In the large-$N$ limit, we find that, for any $\lambda$ at $T=0$ and for $\mathrm{atan}(1/\sqrt{2}) < \lambda \leq \pi/2$ at $T>0$,
the optimal operator is the order parameter of the model, $\hat{O} = \hat{S}_z$, 
while for $0 \leq \lambda \leq \mathrm{atan}(1/\sqrt{2})$ at $T>0$ it is given by the transverse field, $\hat{O} = \hat{S}_x$.

Figure~\ref{fig3} shows the phase diagram of the QFI in the $\cp$-$T$ plane. 
The diagram has the characteristic V-shape illustrated in Fig.~\ref{fig1}.
Figure~\ref{fig3}(a) plots $F_Q[\hat{\rho}_T]/F_Q[\hat{\rho}_0]$.
The crossover temperature $\Tcross(\cp)$ (solid white line) can be identified by the inflection points 
$\partial^2F_Q/\partial{T}^2=0$ and it follows the energy gap $\DeltaE$ (dashed line)
apart a constant multiplication factor, $\DeltaE/\Tcross(\cp) \approx 2.4$. 
Figure~\ref{fig3}(b) plots the logarithmic derivative of the QFI with respect to temperature, 
$\beta \equiv d \log F_Q[\hat{\rho}_T] / d \log T$, giving the scaling exponent for the thermal decay of $F_Q[\hat{\rho}_T]$.
We clearly distinguish regions characterized by constant values of $\beta$ and corresponding to the different plateaus of Fig.~\ref{fig1}:
$\beta = 0$ in the TP and QP, $\beta=-1$ in the CT and, finally, $\beta=-2$ in the MEP. 

These results can be fully understood analytically in the large-$N$ limit via an Holstein-Primakoff approach~\cite{DusuelPRA2004, dusuel2005}.
An expansion in powers of $1/N$ allows to rewrite Eq.~(\ref{HLMG}) as~\cite{JavanainenPRA1999, ShchesnovichPRA2008, Buonsante2012} 
\be \label{HWKB}
\frac{\hat{H}_{\rm BJJ}}{\Eunit} = \frac{N}{2}\sin\cp \left[-\frac{2}{N^2} \frac{\partial}{\partial z} \sqrt{1-z^2} \frac{\partial}{\partial z} + V_\cp(z) \right].
\ee
Here, $z = (N_a-N_b)/N \in[-1,1]$ where $N_{a,b}$ is the number of particles in the mode $\ket{a}$ and $\ket{b}$, respectively. 
$V_\cp(z)=-\frac{z^2}{2} \cot\cp -\sqrt{1-z^2}$ is an effective Ginzburg-Landau potential~\cite{TrenkwalderNATPHYS2016}, 
whose profile has a major role in determining the ground-state structure.
Due to the term $N^{-2}$ in the kinetic energy, the ground state and low-energy excited states are sharply localized 
around the minima $z_0$ of the potential $V_\lambda(z)$.
Thus, for large $N$ we can Taylor expand the Hamiltonian~(\ref{HWKB}) around $z_0$ and retain only the quadratic terms in $z-z_0$.

\begin{figure}[t!]
\includegraphics[width=\columnwidth]{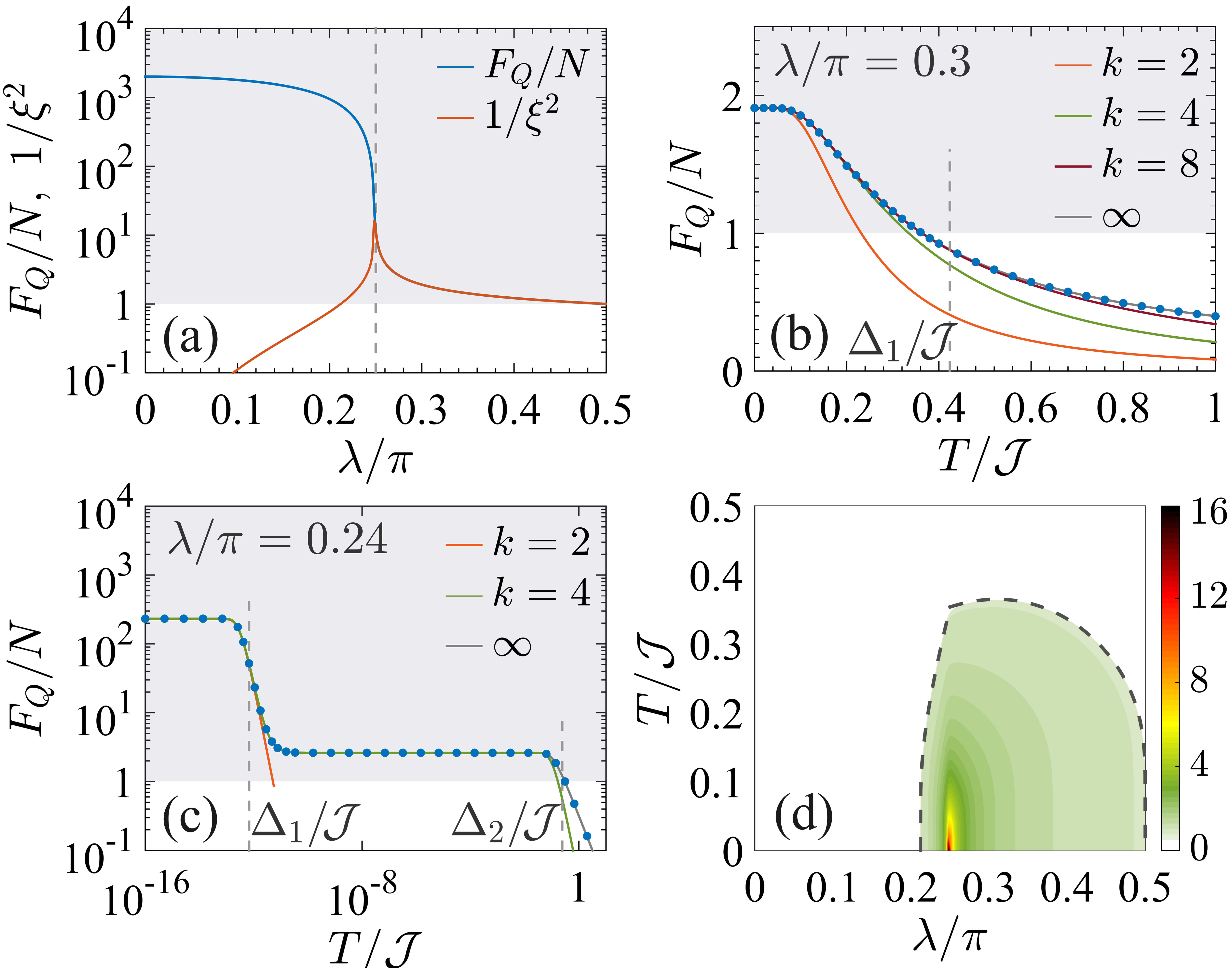}
\caption{
{\bf QFI for the BJJ model.}
(a) Fisher density $F_Q[\ket{\psi_0}]/N$ (blue line) and inverse spin-squeezing parameter 
$1/\xi^2$ (orange line) as a function of $\cp$ for the ground state of Eq.~(\ref{HLMG}). 
The two lines superpose for $\lambda \gtrsim \cpc$.
The vertical dashed line signals the critical point $\cpc = \pi/4$.
Panels (b) and (c) show the Fisher density $F_Q[\hat{\rho}_T]/N$ (dots) as a function of $T$ 
for (b) $\cp = 0.3 \pi >\cpc$ and (c) $\cp = 0.24 \pi<\cpc$.
Solid lines are analytical curves, Eqs.~(\ref{QFImodes}) and (\ref{ferroQFIkmodes}), for different values of the cutoff $k$. 
The vertical dashed lines indicates $T=\Delta_{1,2}$. 
In panels (a)-(c) the shaded area indicates multipartite entanglement $F_Q/N>1$. 
(d) Fisher density $F_Q[\hat{\rho}_T]/N$ (color scale) in the $\cp$-$T$ phase diagram. 
Multipartite entanglement is witnessed at nonzero temperature in the colored region, where $F_Q[\hat{\rho}_T]/N>1$, 
while $F_Q[\hat{\rho}_T]/N \leq1$ corresponds to the white region.
The dashed line is the analytical boundary of $F_Q[\hat{\rho}_T]=N$ in the thermodynamic limit, given by Eq.~(\ref{paraTcross})
for $\lambda > \cpc$ and Eq.~(\ref{ferroTcross}) for $\lambda < \cpc$. 
In all the panels, numerical data are obtained for $N=2000$.} 
\label{fig4}
\end{figure}

{\it Paramagnetic phase, $\cp>\cpc$.} In this case, Eq.~(\ref{HWKB}) reduces to the Hamiltonian of an effective harmonic oscillator centered at $z_0=0$,
\be \label{Hharmonic} 
\frac{\hat{H}_{\rm BJJ}}{\Eunit} = \frac{N}{2}\sin\cp \left[ -\frac{2}{N^2} \frac{\partial^2}{\partial z^2} + \frac{1-\cot\cp}{2}z^2 \right],
\ee
with gap $\Delta_1 = E_1 - E_0 = \Eunit\sqrt{1-\cot\cp}\sin\cp$ in the particle spectrum.
Equation (\ref{Hharmonic}) provides a careful description of the system for energies and temperatures $E_n,T \ll \Eunit N \sin\cp (2-\cot\cp)/4$,
such that the populated eigenstates are only those strongly localized around $z_0=0$
and negligible at the boundaries $z\approx\pm1$. 

At $T = 0$, only the ground state of the harmonic oscillator is populated and 
$F_Q[\ket{\psi_0}] = 4 (\Delta \hat{S}_z)^2 = N/\sqrt{1-\cot\cp}$~\cite{PezzeRMP, MaPRA2009}.
Notice that this variance diverges in the limit $\lambda\to\cpc$
where the potential $V_\cp(z)$ is no longer harmonic.
The QFI is extensive: it linearly grows with the system size $N$.
In particular, at $\cp=\pi/2$ we have $F_Q[\ket{\psi_0}] = N$, 
consistently with the fact that the ground state is separable and given by the coherent spin-polarized state
$(\ket{a} + \ket{b})^{\otimes N}/2^{N/2}$.
Furthermore, we have squeezing of the spin fluctuation 
$(\Delta\hat{S}_y)^2 = N \sqrt{1-\cot \lambda} < N$ below the projection-noise limit
and sufficiently high $\langle\hat{S}_x\rangle$ such that the state is also spin squeezed 
$\xi^2 = N(\Delta\hat{S}_y)^2/\langle\hat{S}_x\rangle^2 = \sqrt{1-\cot \cp}< 1$~\cite{PezzeRMP, MaPRA2009}, 
where $\xi^2$ is the Wineland spin-squeezing parameter~\cite{wineland1994, ma2011}.
The QFI and the spin-squeezing parameter at zero temperature are shown in Fig.~\ref{fig4}(a).

At finite temperature we calculate $F_Q[\hat{\rho}_T]$
taking into account the first $k$ harmonic oscillator modes
and using $\vert \bra{\psi_n} \hat{S}_z \ket{\psi_m} \vert^2 = \sigma^2 [n\delta_{n,m+1} + (n+1)\delta_{n,m-1}]$ with
$\sigma^2 = \frac{N}{4}(1-\cot\cp)^{-1/2}$: 
\be \label{QFImodes}
\frac{F_Q[\hat{\rho}_T]}{F_Q[\ket{\psi_0}]} =  \tanh \bigg( \frac{\Delta_1}{2T} \bigg) \times 
\bigg( 1 - k \frac{\neper^{\Delta_1/T}-1}{\neper^{k\Delta_1/T}-1} \bigg).
\ee
For $k=2$ we recover the two-mode approximation leading to Eq.~(\ref{QFIfactorization}), 
which agrees with the case $k>2$ when $\neper^{-\Delta_1/T} \ll 1$ \cite{NOTA03}. 
A calculation of Eq.~(\ref{QFImodes}) for $k \to \infty$ gives
\be \label{paraQFIfull}
\frac{F_Q[\hat{\rho}_T]}{F_Q[\ket{\psi_0}]}  = \tanh \bigg( \frac{\Delta_1}{2T} \bigg),
\ee
which is in very good agreement with the numerical results of $F_Q[\hat{\rho}_T] $ in the temperature range of interest, as shown in Fig.~\ref{fig4}(b)~\cite{NOTA04}.
For $T \gg \Delta_1$ we find $F_Q[\hat{\rho}_T] \sim 1/T$ at any value of $\lambda$. 
However, Eq.~(\ref{paraQFIfull}) is valid for all $T$ only in the thermodynamic limit: 
for a finite system size, at $T \gg \max_n E_n \propto N$ we recover the MEP regime, where $F_Q[\hat{\rho}_T] \sim 1/T^2$.

{\it Criticality, $\cp=\cpc$.} At $T=0$ the QFI is superextensive: it scales more rapidly than the system size.
A scaling ansatz, with critical exponents $\Delta_Q=1/3$ and $z=1/3$~\cite{HaukeNATPHYS2016}, reveals that
$F_Q/N \sim N^{1/3}$ as a function of $N$~\cite{HaukeNATPHYS2016, MaPRA2009, PezzeRMP}, which is confirmed by numerical calculations.
Notice also that the spin squeezing is $\xi^{-2} \sim F_Q/N $~\cite{DusuelPRA2004, Frerot2017}.
We recall that the energy gap $\Delta_1$ vanishes as $N^{-z}$.  
At finite temperature $T \gg \Delta_1$, we have $F_Q/N \sim T^{-1}$ according to Eq.~(\ref{QC}).

{\it Ferromagnetic phase, $\cp<\cpc$.} 
For sufficiently large $N$, we can calculate the QFI using the semiclassical model of Eq.~(\ref{HWKB}).
The effective potential $V_\lambda(z)$ has a symmetric double-well profile
with two minima located in $z_0=\pm\sqrt{1-\tan^2\cp}$ \cite{ShchesnovichPRA2008}.
Equation~(\ref{HWKB}) takes the form
\be \label{Hdoubleharmonic}
\frac{\hat{H}_{\rm BJJ}}{\Eunit} = \frac{N}{2}\frac{\sin^2\cp}{\cos\cp} \left[ -\frac{2}{N^2} \frac{\partial^2}{\partial z^2} + \frac{1-\tan^2\cp}{2\tan^4\cp} \left(z-z_0\right)^2 \right]
\ee
when locally approximating each well as a harmonic oscillator.

At $T=0$ the QFI, calculated for the ground state of Eq.~(\ref{Hdoubleharmonic}), 
is superextensive, $F_Q[\ket{\psi_0}]/N = N \left( 1-\tan^2\cp \right)$~\cite{ShchesnovichPRA2008, PezzeRMP}, 
see Fig.~\ref{fig4}(a).
In particular, in the limit $\cp\to0$, the ground state is the NOON state
$(\ket{a}^{\otimes N} + \ket{b}^{\otimes N})/\sqrt{2}$, 
given by a coherent symmetric superposition of $N$ particles in mode $\ket{a}$ and $\ket{b}$, 
that has the highest possible value of the QFI,
$F_Q[\ket{\psi_0}]/N=N$~\cite{PezzePRL2009}.

At $T>0$ it is important to distinguish the case of finite $N$, 
where the energy gap $\Delta_1 \propto \exp(- N |1-\cot\cp|^{4/3})$ damps exponentially, 
and the thermodynamic limit $N \to \infty$, where $\Delta_1 = 0$.
In the latter case, the ground state is doubly-degenerate ($\mu=2$)
and  separated from the doubly-degenerate first excited state ($\nu=2$) by the energy
gap $\Delta_2 = E_2 - E_1 = \Eunit \sqrt{1-\tan^2\cp}\cos\cp$.
For arbitrary small but finite temperatures, $0< T\ll \Delta_2$, 
the system is described by the incoherent mixture $\hat{\rho}_0 = ( \ket{\psi_0}\bra{\psi_0} + \ket{\psi_1}\bra{\psi_1} )/2$.
Its QFI is  
\be \label{ferroQFIplateau}
F_Q[\hat{\rho}_0] = N \times \left\{
\begin{array}{ll} 
\sqrt{1-\tan^2\cp} & {\rm for} \ 0\leq\lambda\leq\lambdadir \\
\displaystyle\frac{\tan^2\cp}{\sqrt{1-\tan^2\cp}} & {\rm for} \ \lambdadir<\lambda<\cpc
\end{array}
\right. \, , 
\ee
where $\lambdadir=\mathrm{atan}(1/\sqrt{2})$ arises from the optimization of the operator: 
$\hat{O}=\hat{S}_x$ for $\lambda\leq\lambdadir$, while $\hat{O}=\hat{S}_z$ for $\lambda>\lambdadir$.
We can calculate the QFI using a $k$-mode approximation 
({\it i.e.}~taking the first $k$ states in each harmonic well). 
By means of $\vert \bra{\psi_n} \hat{S}_z \ket{\psi_m} \vert^2 = \frac{1-(-1)^{n+m}}{2}\{z_0^2\delta_{\tilde{n},\tilde{m}}+\sigma^2 [\tilde{n}\delta_{\tilde{n},\tilde{m}+1} + (\tilde{n}+1)\delta_{\tilde{n},\tilde{m}-1}]\}$, 
where $\tilde{n}=\lfloor\frac{n}{2}\rfloor$ and $\sigma^2=\frac{N}{4}\tan^2\cp/\sqrt{1-\tan^2\cp}$,
we can evaluate the QFI in Eq.~(\ref{QFImixed1}), obtaining
\be \label{ferroQFIkmodes}
\frac{F_Q[\hat{\rho}_T]}{F_Q[\hat{\rho}_0]} = \tanh \bigg( \frac{\Delta_2}{2T} \bigg) 
\times 
\bigg( 1 - \frac{k}{2} \frac{\neper^{\Delta_2/T}-1}{\neper^{k\Delta_2/2T}-1} \bigg),
\ee
that becomes 
\be \label{ferroQFIfull}
\frac{F_Q[\hat{\rho}_T]}{F_Q[\hat{\rho}_0]} = \tanh \left( \frac{\Delta_2}{2T} \right),
\ee
when taking the limit $k \to \infty$. 
It should be noticed that Eq.~(\ref{ferroQFIplateau}) is a factor $N$ smaller than the zero-temperature value $F_Q[\ket{\psi_0}]$.
For large but finite $N$, the rapid decay from $F_Q[\ket{\psi_0}]$ to $F_Q[\hat{\rho}_0]$ is described by the decaying function $\tanh^2(\Delta_1/2T)$ [see Fig.~\ref{fig4}(c)], 
as predicted by Eq.~(\ref{QFIfactorization}) for a purely two-mode approximation.
In the thermodynamic limit we have that $\Delta_1\to 0$ and we thus find a discontinuous jump of the QFI from its $T=0$ value and
the plateau described by Eq.~(\ref{ferroQFIfull}).  
This behavior characterizes the TP of Fig.~\ref{fig1}.

The QFI in the different regimes is illustrated in Fig.~\ref{fig4}(c), where we show a very good agreement between the analytical predictions and the 
numerical results. 
Also in this case, for large enough temperature $T \gg \Delta_2$ 
we recover $F_Q[\hat{\rho}_T] \sim 1/T$ as the leading term in the Taylor expansion of the $\tanh$ function.

\begin{figure*}[t!]
\includegraphics[width=\textwidth]{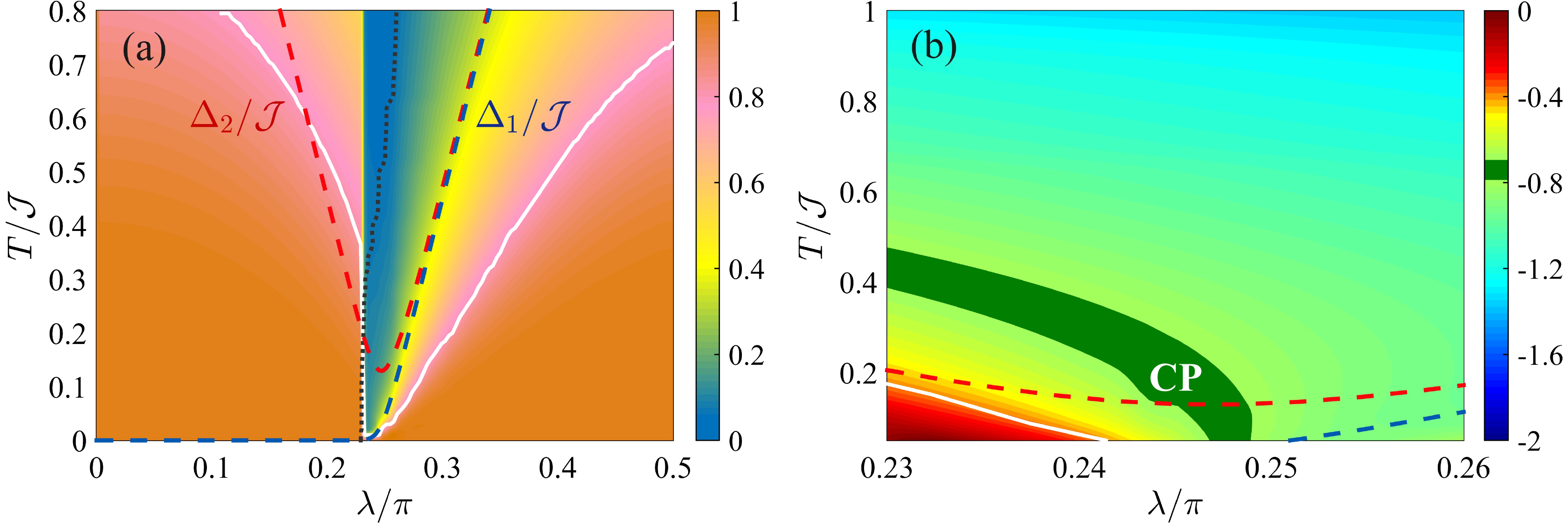} \\
\caption{{\bf Phase diagram of the Ising model in transverse field.}
(a) QFI normalized to its low-temperature value, $F_Q[\hat{\rho}_T]/F_Q[\hat{\rho}_0]$ (color scale), in the $\cp$-$T$ phase diagram. 
(b) Scaling coefficient $\beta = d \log F_Q[\hat{\rho}_T] / d \log T$ in the vicinity of the critical point. 
In both panels, the white solid line is the crossover temperature $\Tcross(\cp)$.
The blue and red dashed lines indicate $\Delta_1$ and $\Delta_2$, respectively.
In both panels, $N=50$.} 
\label{fig5}
\end{figure*}

{\it Multipartite entanglement.}
In Fig.~\ref{fig4}(d) we plot $F_Q[\hat{\rho}_T] / N$ in the $\lambda$-$T$ plane.
Multipartite entanglement witnessed by the QFI is found in the colored region that is bounded by the separability condition $F_Q[\hat{\rho}_T] / N = 1$.
It is worth pointing out that multipartite entanglement is considered here among distinguishable spin-1/2 particles restricted to occupy
permutationally symmetric quantum states. 
In practical realizations, such as a Bose-Einstein condensate in double-well trap, these spin-1/2 particles are not addressable.
In the limit $N \to \infty$, $\kappa$-partite entanglement witnessed by the QFI is found at temperatures  
\be \label{paraTcross}
T < \frac{\Eunit \sqrt{1-\cot\cp}\sin\cp}{2 \, {\rm atanh} \left(\kappa\sqrt{1-\cot\cp}\right)}
\ee
for $\lambda > \cpc$, as obtained from Eq.~(\ref{paraQFIfull}), and at 
\be \label{ferroTcross}
T <  \frac{\Eunit \sqrt{\cot^2\cp-1}\sin\cp}{2 \, {\rm atanh}\left(\kappa\cot\cp\sqrt{\cot^2\cp-1}\right)}
\ee
for $\lambda < \cpc$, following Eq.~(\ref{ferroQFIfull}).
Equations (\ref{paraTcross}) and (\ref{ferroTcross}) are shown as dashed lines in Fig.~\ref{fig4}(d).
In particular, as noticed above, multipartite entanglement in the ground state of the ferromegnetic phase is extremely fragile 
to temperature. 
Moreover, in the thermodynamic limit, we find that no entanglement is witnessed by the QFI at $T>0$ for $\cp\leq\evap$, where 
\be
\cot\cp^* = \sqrt{\frac{1+\sqrt{5}}{2}} + \mathcal{O}\left(N^{-1}\right).
\ee
Remarkably, at finite temperature, the QFI detects the same amount of entanglement detected by the spin-squeezing parameter: 
$1/\xi^2=F_Q/N$ for $T>0$ in the limit $N\gg1$:
thermal noise is responsible for a loss of coherence entailing a spread of spin fluctuations in any direction.
In particular, $\evap$ is the point at which the spin squeezing ceases to detect entanglement ($\xi^2=1$) even at $T=0$ 
because of the vanishing $\langle\hat{S}_x\rangle$.
When maximizing Eqs.~(\ref{paraTcross}) and (\ref{ferroTcross}) over $\lambda$, we obtain that entanglement 
detected by the QFI survives up to $T/\Eunit\approx0.4$, see Fig.~\ref{fig2}(d).

\subsection{One-dimensional Ising model in a transverse field} 
\label{SubSectionIsing}

The one-dimensional quantum Ising chain in a transverse field~\cite{PfeutyANNPHYS1970,Dutta2015},
\be \label{Hising} 
\frac{\hat{H}_{{\rm TFI}}}{\Eunit} = - \cos\cp \sum_{i=1}^{N-1} \hat{\sigma}_z^{(i)}\hat{\sigma}_{z}^{(i+1)} + \sin\cp\sum_{i=1}^N \hat{\sigma}_x^{(i)},
\ee
describes $N$ distinguishable spin-$1/2$ particles interacting via a nearest-neighbor exchange energy $\Eunit\sin\cp$ (open boundaries are assumed)
and subject to a transverse magnetic field of strength $\Eunit\cos\cp$, with $\cp\in[0,\pi/2]$.
The interaction term favors ferromagnetic ordering (with all spins aligned along $\pm z$), while the 
transverse field favors polarization (with all spins aligned along $-x$).
In the thermodynamic limit $N\to \infty$ and for $T=0$, Eq.~(\ref{Hising}) exhibits a QPT at $\cpc=\pi/4$
between a paramagnetic phase (for $\cpc<\cp \leq \pi/2$) 
and a ferromagnetic phase (for $0 \leq \cp < \cpc$). 
The Ising model in a transverse field is a testbed of quantum criticality~\cite{Sachdev2011}. 

{\it Phase diagram.}
Figure~\ref{fig5} shows the phase diagram of the QFI in the  $\cp$-$T$ plane, 
where the QFI is optimized with respect to the collective operator $\hat{O} = \frac{1}{2} \sum_{i=1}^N \hat{\sigma}_{x,y,z}$.
In particular, the black line in Fig.~\ref{fig5}(a) marks a region where the optimal operator is 
$\hat{O} = \frac{1}{2} \sum_{i=1}^N \hat{\sigma}_{x}$ (on the left side of the line) and the one where the optimal operator is  
the order parameter of the transition, $\hat{O} = \frac{1}{2} \sum_{i=1}^N \hat{\sigma}_z$ (on the right side of the line).
The diagram displays the characteristic V-shaped structure radiating from the critical point, as in Fig.~\ref{fig1}.
In Fig.~\ref{fig5}(a) we can recognize the CP (for $\lambda > \cpc$) and the TP (for $\lambda < \cpc$).
Therein, the QFI $F_Q[\hat{\rho}_T]$ is approximatively constant as a function of temperature and equal to 
its low-temperature value $F_Q[\hat{\rho}_0]$ -- we recall that $\hat{\rho}_0$ is given by the ground state $\ket{\psi_0}$ 
in the CP and by the incoherent superposition of the two lowest energy eigenstates in the TP. 
We also see that $\Tcross$ (solid white line) follows the energy gap $\Delta$ (dashed line). 
The finite jump discontinuity of $\Tcross$ that is visible in the figure is due to the sudden change
of optimal operator $\hat{O}$ and prominently manifests only for small $N$.
In Fig.~\ref{fig5}(b) we plot the logarithmic derivative $\beta = d \log F_Q[\hat{\rho}_T] / d \log T$ in the vicinity of the critical point, 
which provides the scaling of the QFI with temperature. 
According to the scaling ansatz (see Sec.~\ref{R&M}), 
using $\Delta_Q = 3/4$ and $z=1$~\cite{HaukeNATPHYS2016}, we find $F_Q[\hat{\rho}_T] \propto T^{-3/4}$ in the QP.
Numerical results are plagued by finite size effects and we do not observe a clear plateau for $\beta$.
We argue (supported by a finite-size study, yet limited to $N \lesssim 100$) that the CP, where $\beta=-0.75$, 
approximatively coincides with the green region in the figure, which highlights values of $\beta \in [-0.78,-0.72]$.

\begin{figure}[t!]
\includegraphics[width=\columnwidth]{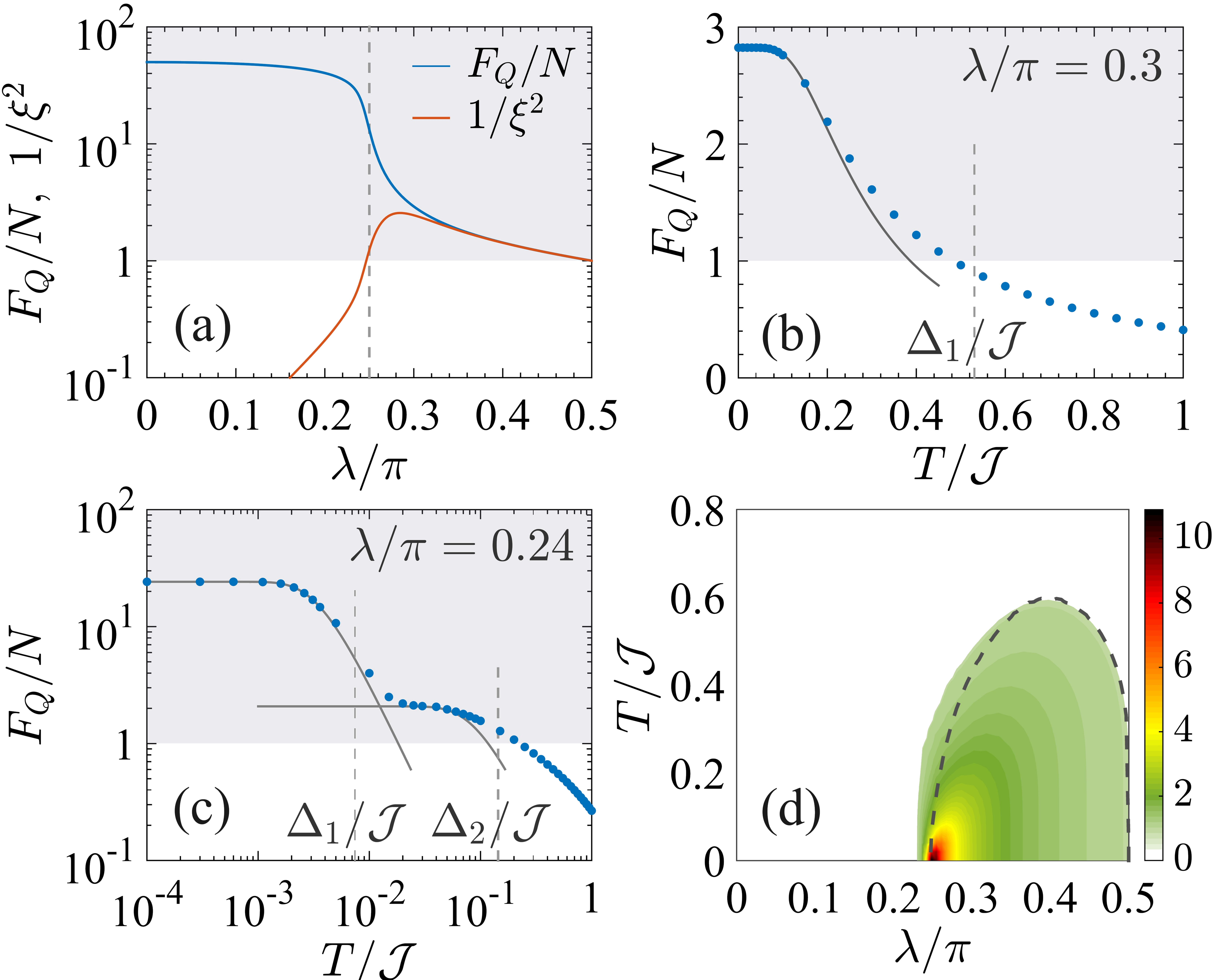} \\
\caption{{\bf QFI for the Ising model in transverse field.}
(a) Fisher density $F_Q[\ket{\psi_0}]/N$ (blue line) and inverse spin squeezing (orange line) 
for the ground state of Eq.~(\ref{Hising}) as a function of $\cp$. 
The vertical dashed line signals the critical point $\cpc$.
Panels (b) and (c) show the typical decay of the Fisher density $F_Q[\hat{\rho}_T]/N$ as a function of $T$ 
in the paramagnetic (b) and ferromagnetic (c) phase. 
The solid lines are $\tanh^2(\Delta_1/2T)$ or $\tanh^2(\Delta_2/2T)$.
In panels (a)-(c) the shaded area indicates multipartite entanglement. 
(d) Fisher density $F_Q[\hat{\rho}_T]/N$ (color scale) in the $\cp$-$T$ phase diagram.
The dashed line is the spin-squeezing boundary $\xi^2=1$.
In all panels $N=50$. } 
\label{fig6}
\end{figure}

The behavior of the QFI is further inspected in Fig.~\ref{fig6}.
In panel (a) we show the QFI (blue line) for the ground state of Eq.~(\ref{Hising}) and compare it to the spin squeezing
$\xi^2 = N (\Delta \hat{S}_y)^2/ \mean{\hat{S}_x}^2$ (orange line).
Similarly to the BJJ model, for $\lambda>\cpc$ the QFI is larger than $N$ -- signaling multipartite entanglement -- but extensive, 
{\it i.e.}~the Fisher density $F_Q[\ket{\psi_0}]/N$ does not scale with $N$ \cite{LiuJPA2013}.
For $\lambda \leq \cpc$, the QFI is superextensive.
It scales as $F_Q[\ket{\psi_0}]/N \sim N$ for $\lambda < \cpc$~\cite{LiuJPA2013} and 
as $F_Q[\ket{\psi_0}]/N \sim N^{3/4}$ at $\lambda = \cpc$~\cite{HaukeNATPHYS2016}.
While the spin squeezing agrees with the QFI close to $\lambda=\pi/2$, it sharply decays at $\cpc$.
Indeed, a numerical study as a function of $N$ (up to $N=500$) for $\lambda=\cpc$ reveals that  
$\xi^2=1$ and in particular it does not scale with $N$, see also \cite{LiuJPA2013, Frerot2017}.
This is in sharp contrast to the results of the BJJ model where the QFI and the spin-squeezing parameters 
for the ground state have the same scaling at the critical point, see Fig.~\ref{fig4}(a).
The typical behavior of $F_Q[\hat{\rho}_T]$ as a function of temperature for $\cp>\cpc$ is shown in Fig.~\ref{fig6}(b).
The solid line is $F_Q[\hat{\rho}_T] / F_Q[\hat{\rho}_0] = \tanh^2(\Delta_1/2T)$, namely Eq.~(\ref{QFIfactorization}) with $\mu = \nu = 1$.
For $\cp<\cpc$, the ground state becomes doubly degenerate ($\mu=2$)
in the thermodynamic limit $N \to \infty$, and so does the first excited state.
For finite $N$ (the gap $\DeltaE_1$ being exponentially small \cite{NOTA05}),
the behavior of $F_Q[\hat{\rho}_T]$ as a function of temperature is shown in Fig.~\ref{fig6}(c): 
the decay from the zero-temperature value occurs around a finite (exponentially small in $N$) temperature $T\lesssim\Delta_1$, 
while a slower decay takes place for $T\gg\Delta_2$.
The constant plateau $F_Q[\hat{\rho}_T] \approx F_Q[\hat{\rho}_0]$ found for $\Delta_1 \ll T\ll \Delta_2$ defines the TP.

{\it Multipartite entanglement.}
The multipartite entanglement between spin-1/2 particles detected by the QFI survives at finite temperature in the colored region of Fig.~\ref{fig6}(d),
bounded by $F_Q[\hat{\rho}_T]/N=1$.
This region fans out from the zero-temperature noninteracting $\cp=\pi/2$ corner, 
where the ground state of Eq.~(\ref{Hising}) is separable and $F_Q[\hat{\rho}_0]/N=1$.
We can compare the condition $F_Q[\hat{\rho}_T]/N=1$ with the spin-squeezing coefficient $\xi^2=1$ (dashed line)~\cite{Frerot2017}.
The loss of spin squeezing follows the loss of thermal entanglement only for $\lambda \approx \pi/2$, 
while around $\cpc$ we have entangled states recognized by the QFI that are not spin squeezed.
Furthermore, the multipartite entanglement region in Fig.~\ref{fig6}(d)
reaches a maximum extension $T/\Eunit\approx0.6$ at $\cp=0.4\pi$.
Notice that this threshold temperature is higher than the one for the BJJ model. 
Interestingly, the threshold $T/\Eunit\approx0.3$ at $\cp= \cpc$ is consistent with 
the temperature up to which other thermal signatures of criticality persists~\cite{KoppNATPHYS2005, KinrossPRX2014}.
Finally, for $\cp \lesssim 0.2 \pi$ multipartite entanglement is no more witnessed by the QFI at any $T>0$.

\begin{figure*}[t!]
\includegraphics[width=\textwidth]{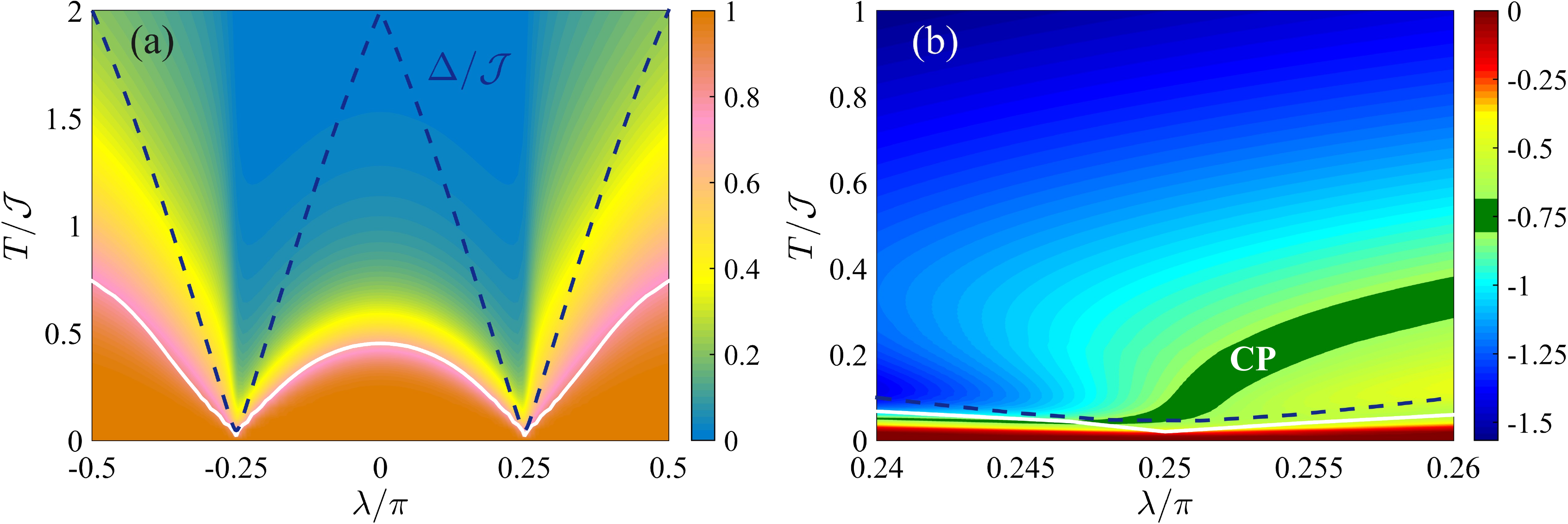}
\caption{
{\bf Phase diagram of the Kitaev chain with short-range pairing.}
(a) QFI normalized to its low-temperature value, $F_Q[\hat{\rho}_T]/F_Q[\ket{\psi_0}]$
(color scale),
in the $\lambda$-$T$ phase diagram.
(b) Scaling coefficient $\beta = d \log F_Q[\hat{\rho}_T] / d \log T$
as a function of $\cp$ and $T$.
In both panels the white line is $\Tcross$ and the blue dashed line is the energy gap $\Delta$.
Here $N=50$ and $\alpha=100$.
} 
\label{fig7}
\end{figure*}

\section{Applications: topological QPTs}  
\label{SectionTopological} 

In the following we study the one-dimensional Kitaev model~\cite{KitaevPU2001, AliceaRPP2012} 
for spinless fermions hopping in a tight-binding lattice with p-wave superconducting pairing. 
With respect to the original model~\cite{KitaevPU2001}, we consider variable range for the pairing \cite{VodolaPRL2014, VodolaNJP2016}. 
The Hamiltonian is
\beq \label{HKitaev0} 
\hat{H}_{\rm K} = & - & \displaystyle \frac{J}{2} \sum_{i=1}^\sites \big( \hat{a}^\dagger_i \hat{a}_{i+1} + {\rm H.c.} \big) 
- \mu \sum_{i=1}^\sites \Big(\hat{n}_i - \frac{1}{2}\Big) \nonumber \\ 
& + & \displaystyle \frac{\Omega}{2} \sum_{i=1}^\sites \,\sum_{\ell=1}^{\sites-i} d_\ell^{-\alpha} 
\big( \hat{a}_i \hat{a}_{i+\ell} + {\rm H.c.} \big),
\eeq
where $\hat{a}^\dagger_i$ is a fermionic creation operator at $i$-th site 
(satisfying the anticommutation relation $\{\hat{a}_i,\hat{a}_j^\dagger\}=\delta_{i,j}$) and 
$\hat{n}_i=\hat{a}_i^\dagger\hat{a}_i$ counts the number of fermions in the $i$-th site.
The amplitude of the hopping between different lattice sites is $J$
and the chemical potential of the chain is $\mu$.
The superconducting pairing has strength $\Omega$ and
range specified by $d_\ell^\alpha$, where $d_\ell$ is a site-to-site distance and $\alpha>0$: 
$\alpha\to\infty$ corresponds to nearest-neighbor pairing, while $\alpha=0$ accounts for infinite-range pairing.
For a closed ring, $d_\ell=\ell$ ($d_\ell=\sites-\ell$) if $\ell\leq\sites/2$ ($\ell>\sites/2$).
In Eq.~(\ref{HKitaev0}) we consider antiperiodic boundary conditions ($\hat{a}_{N+1}=-\hat{a}_1$).

The Hamiltonian~(\ref{HKitaev0}) can be exactly diagonalized 
by a Bogoliubov transformation~\cite{Lieb1961} for any $\alpha$. 
The quasiparticle spectrum reads~\cite{VodolaPRL2014}
\be \label{spect}
\epsilon_{k} = \sqrt{ \big(J\cos k + \mu\big)^2 + \big(\Omega f_\alpha(k)/2\big)^2},
\ee
where $k = 2\pi(n+\frac{1}{2})/\sites$ are the quasimomenta of the excitations ($ n=0,1,...,\sites-1$) 
and $f_\alpha(k) = \sum_{\ell=1}^{\sites-1} \sin(k\ell)/d_\ell^\alpha$.
The first energy gap in the many-body spectrum $\Delta=\min_k\epsilon_k$ 
corresponds to the energy necessary to create one elementary excitation. 
The ground state of the Kitaev chain Eq.~(\ref{HKitaev0}) reads 
$\ket{\psi_0} = \prod_{0\,<\,k\,<\,\pi} (\cos\frac{\theta_k}{2} - \ii\,\sin\frac{\theta_k}{2} \, \hat{a}_k^\dag\hat{a}_{-k}^{\dag}) \ket{0}$,
with $\sin\theta_k=-\Omega f_\alpha(k)/(2\epsilon_k)$, $\cos\theta_k=-(J\cos{k}+\mu)/\epsilon_k$, 
$\hat{a}_k^\dag = \frac{1}{\sqrt{N}} \sum_{i=1}^N e^{-\ii k i} \hat{a}_i^\dag$ 
being the Fourier transform of $\hat{a}_{i}^\dag$ 
and $\ket{0}$ denoting the vacuum of quasiparticles.
The ground state hosts different topological phases that can be characterized 
by the winding number $W=\frac{1}{2\pi}\int_0^{2\pi}\frac{\ud \theta_k}{\ud k} \ud k$.
At zero temperature, the mean number of fermions in the system is given by 
$\sum_i\langle\hat{n}_i\rangle = \sum_k \sin^2\frac{\theta_k}{2} \leq N$.

Here, we study the QFI calculated with respect to the nonlocal operators ($\varrho=x,y$)
\be \label{KitaveNonlocalOperator}
\hat{O}_{\varrho}^{(\pm)} = \sum_{i=1}^\sites (\pm1)^i \frac{\hat{a}_i^\dag \neper^{\ii\pi\sum_{j=1}^{i-1}\hat{n}_j} 
+ (-1)^{\delta_{\varrho y}} \neper^{-\ii\pi\sum_{j=1}^{i-1}\hat{n}_j} \hat{a}_i}{2\ii^{\delta_{\varrho y}}} 
\ee 
and the local operators
\be \label{KitaveLocalOperator}
\hat{O}_z^{(\pm)} = \frac{1}{2} \sum_{i=1}^N (\pm1)^i  (2\hat{n}_i -1).
\ee
This choice is suggested by the Jordan-Wigner transformation 
$\hat{\sigma}_+^{(i)} = 2 \hat{a}_i^\dag \neper^{\ii \pi \sum_{j=1}^{i-1} \hat{n}_j}$ 
and $\hat{\sigma}_-^{(i)} = 2 \neper^{-\ii \pi \sum_{j=1}^{i-1} \hat{n}_j} \hat{a}_i$, 
that relates fermionic creation and annihilation operators 
to the Pauli ladder operators $\hat{\sigma}_+$ and $\hat{\sigma}_-$ of spin-1/2 particles~\cite{Lieb1961}.
Via Jordan-Wigner trasformation, 
the nearest-neighbor Kitaev chain ($\alpha=\infty$) maps onto the Ising model in a transverse field~\cite{Lieb1961, NOTA06}, and operators in Eqs.~(\ref{KitaveNonlocalOperator}) and (\ref{KitaveLocalOperator}) become the local collective spin operators
$\hat{O}_{\varrho}^{(\pm)} = \frac{1}{2} \sum_{i=1}^\sites (\pm1)^i \hat{\sigma}_{\varrho}^{(i)}$, 
being $\hat{\sigma}_{\varrho}^{(i)} = \big(\hat{\sigma}_+^{(i)} + (-1)^{\delta_{\varrho y}} \hat{\sigma}_-^{(i)}\big)/(2\ii^{\delta_{\varrho y}})$, 
and $\hat{O}_z^{(\pm)} = \tfrac{1}{2}\sum_{i=1}^N (\pm1)^i \hat{\sigma}_z$.
By mean of this transformation, each lattice site maps into an effective spin-1/2 particle: 
the $z$ component of the spin is local in the site (the empty lattice corresponding to spin-down, the filled lattice to spin-up), 
the other $x$ and $y$ components are nonlocal. 
We can then use the bound discussed above~\cite{PezzePRL2009, HyllusPRA2012, TothPRA2012} to witness 
$\kappa$-particle entanglement between the $N$ effective spin-1/2. 
Specifically, $F_Q[\hat{\rho}_T]/\sites > \kappa$ signals $(\kappa+1)$-partite entanglement. 
Recently, operators $\hat{O}_{\varrho}^{(\pm)}$ have been used to demonstrate the superextensivity of the QFI at zero temperature 
in the topological phases of the Kitaev model~(\ref{HKitaev0}), for both short-range and long-range pairing~\cite{PezzePRL2017}.

In the following, we set equal pairing and hopping strengths $\Omega=J$
and take $J=2\Eunit\cos\cp$ and $\mu=2\Eunit\sin\cp$
in order to describe the whole phase diagram through a bounded control parameter $\cp\in [\mathrm{-}\pi/2,\pi/2]$.
The Hamiltonian (\ref{HKitaev0}) thus rewrites as
\beq \label{HKitaev00}
\frac{\hat{H}_{\rm K}}{\mathcal{J}} = & - & \cos \lambda \sum_{i=1}^\sites \big( \hat{a}^\dagger_i \hat{a}_{i+1} + {\rm H.c.} \big) 
- 2 \sin \lambda \sum_{i=1}^\sites \Big(\hat{n}_i - \frac{1}{2}\Big) \nonumber \\ 
& + & \cos \lambda \sum_{i=1}^\sites \,\sum_{\ell=1}^{\sites-i} d_\ell^{-\alpha} 
\big( \hat{a}_i \hat{a}_{i+\ell} + {\rm H.c.} \big).
\eeq

\subsection{Kitaev model with short-range pairing}
 
We consider the case $\alpha=\infty$ where pairing occurs only within nearest-neighbor lattice sites. 
In this case $f_{\alpha}(k) = 2 \sin k$.
As shown by Eq.~(\ref{spect}), the energy gap between the ground state and the first excited state 
vanishes as $\Delta \sim N^{-1}$ in the thermodynamic limit at $\cpc = \pm \pi/4$ (for $k=\pi$ and $k=0$, respectively). 
These quantum critical points separate a topological nontrivial phase with $W=1$ (for $|\lambda| < \pi/4$)
from a trivial phase with $W=0$ (for $|\lambda| > \pi/4$).
This behavior is common for short-range pairing, $\alpha>1$~\cite{PezzePRL2017, VodolaPRL2014}.

{\it Phase diagram.} 
As expected, the results of our study are very similar to the case of the quantum Ising model discussed in Sec.~\ref{SectionSBQPT}.
There is a major difference though: in the Kitaev model the energy gap $\Delta=E_1-E_0$ remains finite for every $\lambda\neq\cpc$, {\it i.e.}~away from the critical points.
Therefore, the system does not host a gapless phase, differently from the ferromagnetic phase of the Ising model.
This is a direct consequence of the fact that the Kitaev model is studied here in the closed chain. 
In the open chain, the Kitaev model hosts a gapless phase for $ |\lambda | < \pi/4$, related to the presence of Majorana edge modes.

\begin{figure}[t!]
\includegraphics[width=\columnwidth]{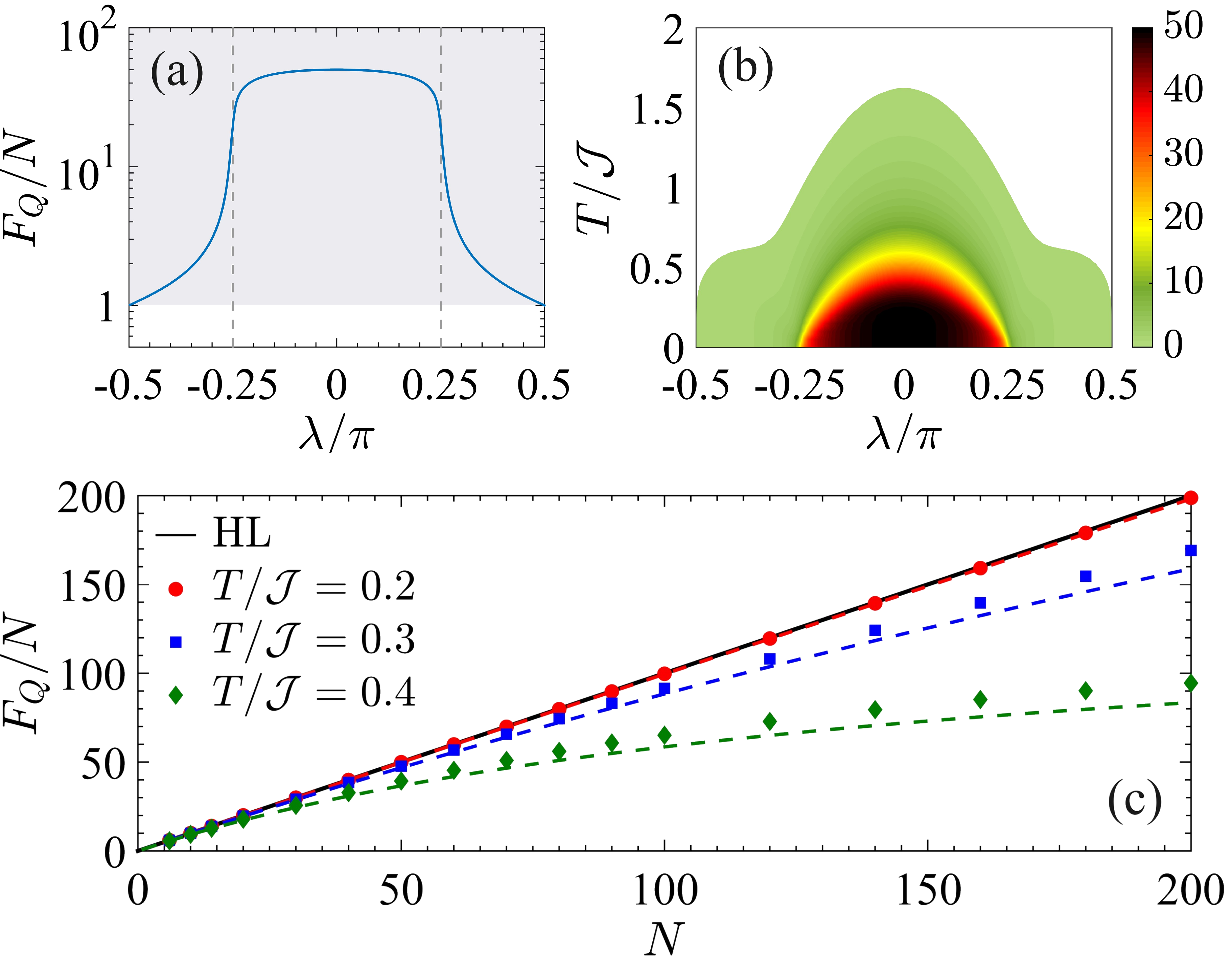}
\caption{
{\bf QFI for the Kitaev chain with short-range pairing.}
(a) Fisher density $F_Q[\ket{\psi_0}]/\sites$ as a function of $\cp$ for the ground state of Eq.~(\ref{HKitaev00}) with $N=50$ and $\alpha=100$. 
The vertical dashed lines signal the critical points $\cpc$. 
The shaded area marks entanglement, $F_Q[\ket{\psi_0}] > \sites$. 
(b) Fisher density $F_Q[\hat{\rho}_T]/\sites$ (color scale) in the $\lambda$-$T$ plane for $N=50$.
The colored area corresponds to $F_Q[\hat{\rho}_T] > \sites$.
(c) Scaling of $F_Q[\hat{\rho}_T]/\sites$ as a function of $N$ for different temperatures. 
The thick black line is the Heisenberg limit $F_Q = \sites^2$, 
the dashed lines are the bound in Eq.~(\ref{KITSRFish}).
} 
\label{fig8}
\end{figure}

\begin{figure*}[t!]
\includegraphics[width=\textwidth]{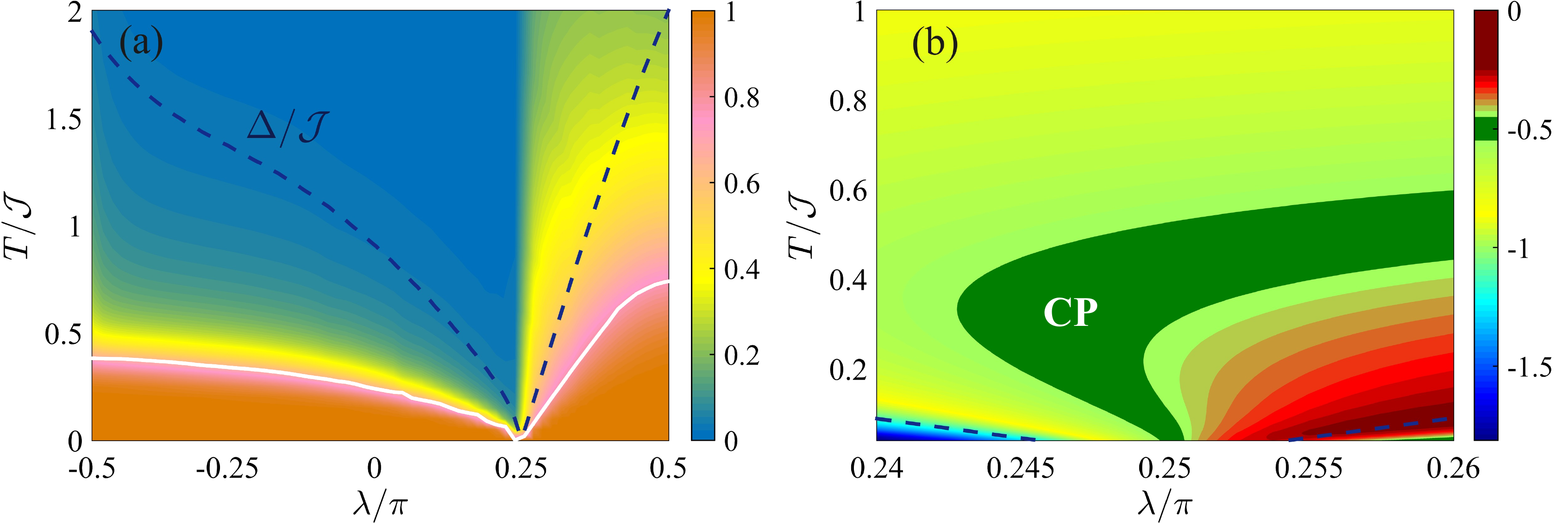} 
\caption{
{\bf Phase diagram of the Kitaev chain with long-range pairing.}
(a) QFI normalized to its low-temperature value, $F_Q[\hat{\rho}_T]/F_Q[\ket{\psi_0}]$
(color scale),
in the $\lambda$-$T$ phase diagram.
(b) Scaling coefficient $\beta = d \log F_Q[\hat{\rho}_T] / d \log T$.
In both panels the white line is $\Tcross$ and the blue dashed line is the energy gap $\Delta$.
Here $N=50$ and $\alpha=0$.
} 
\label{fig9}
\end{figure*}
  
In Fig.~\ref{fig7}(a) we plot the $\lambda$-$T$ phase diagram for $F_Q[\hat{\rho}_T]/F_Q[\ket{\psi_0}]$.
The optimal operator maximizing the QFI is found to be $\hat{O}_x^{(+)}$ for any $\lambda$ and $T$.
We recognize the presence of plateaus at low temperature and the characteristic V-structure around the critical points.
Only QPs are present, due to the nondegenerate nature of the ground state.
The phase diagram is invariant under change of sign of the chemical potential $\lambda\to-\lambda$, 
as expected from the particle-hole symmetry of the Hamiltonian~\cite{AliceaRPP2012}.
The crossover temperature $\Tcross(\cp)$ (solid white line) follows the energy gap (dashed line) 
for $\vert\lambda\vert > \pi/4$, with $\DeltaE/\Tcross\approx2.7$. 
In the region $\vert\lambda\vert < \pi/4$, $\Tcross$ is instead smoothed, 
due to the quasi-degeneracy of the excited states.
In Fig.~\ref{fig7}(b) we plot the logarithmic derivative $\beta = d \log F_Q[\hat{\rho}_T] / d \log T$ around the critical point $\lambda=\pi/4$. 
The QPT is characterized by the same critical exponents as the Ising model, $\Delta_Q = 3/4$ and $z=1$,
and we thus expect a thermal decay $F_Q[\hat{\rho}_T]\sim T^{-3/4}$, according to Eq.~(\ref{QC}).
The region where $\beta \in [-0.8,-0.7]$ is highlighted in the figure.

{\it Multipartite entanglement.}
Figure~\ref{fig8} illustrates the multipartite entanglement witnessed by the QFI.
Panel (a) shows the QFI of the ground state, $F_Q[\ket{\psi_0}]/\sites$, as a function of $\cp$.
The trivial phase $|\cp|>\pi/4$ is characterized 
by an extensive scaling of the QFI for increasing system size $\sites$.
At $\cp=+\pi/2$ ($\cp=-\pi/2$), we find $F_Q[\ket{\psi_0}] = N$, 
according to the fact that the ground state is a separable state 
of occupied (empty) sites $\ket{\psi_0}=\ket{1}^{\otimes\sites}$ ($\ket{\psi_0}=\ket{0}^{\otimes\sites}$), 
where $\{\ket{n}_i\}$ is the occupation basis and $n\in\{0,1\}$ is the occupation number at the $i$-th site.
Divergence of multipartiteness $F_Q/N \sim N$ is instead observed in the phase with nonzero winding number ($|\cp|<\pi/4$)~\cite{PezzePRL2017}.
In particular, $F_Q/\sites=\sites$  at $\cp=0$.
The QPT at $\cpc$ is signalled by a sudden change in the scaling $F_Q/\sites \sim \sites^{3/4}$, 
that is associated to the specific algebraic asymptotic decay observed for the two-site correlation functions~\cite{PezzePRL2017}.

Figure~\ref{fig8}(b) shows the witnessed multipartite entanglement at finite temperature.
Within the region $\vert\lambda\vert < \pi/4$, superextensive multipartite entanglement in the ground state 
survives at finite temperature. 
This robustness is due to the nondegenerate nature of the ground state for any $\cp\neq\cpc$ and it is in sharp contrast with the ferromagnetic phase of the BJJ and Ising model, 
where a superextensive QFI decays exponentially with $N$ at finite temperature.
In particular, at $\cp=0$, where the first excited state is $N$-fold degenerate, Eq.~(\ref{QFIfactorization}) predicts for low temperature
\be \label{KITSRFish}
\frac{F_Q[\hat{\rho}_T]}{N^2} \geq 
\tanh^2 \left(\frac{\Eunit}{T}\right) \frac{1+\neper^{-2\Eunit/T}}{1+N\neper^{-2\Eunit/T}},
\ee
where we have used $F_Q[\ket{\psi_0}] = N^2$, $\Delta=2\Eunit$, $\mu=1$ and $\nu=N$.
For large $N$ the right-hand side of Eq.~(\ref{KITSRFish}) can be well approximated by 
$1/(1+N\neper^{-2\Eunit/T})$ that shows a plateau up to temperatures $T/\Eunit \approx 2/\log N$.
The thermal decay of the QFI is at most logarithmic in $N$. 
This behavior is confirmed in Fig.~\ref{fig8}(c) where, for a fixed temperature, we plot 
$F_Q/N$ as a function of $N$. 
We see that the Heisenberg scaling $F_Q/N \sim N$ survives at finite temperature up to $N\ll\neper^{2\Eunit/T}$.
For larger system size, temperature is responsible for a softening of the power-law scaling.
It is worth noticing that this effect is not related to a vanishing gap, as in the ferromagnetic phase of the Ising model, 
but it is rather due to the diverging degeneracy of the first excited state. 
As emphasized in Eq.~(\ref{QFIfactorization}), the robustness of the QFI to temperature depends indeed on this degeneracy.

\subsection{Kitaev model with long-range pairing}

We study the Kiteav model with $\alpha=0$ where pairing involves fermions in arbitrarily-distant sites. 
In this case $f_{\alpha}(k) = \cot(k/2)$, which diverges at $k=0$.
The energy gap vanishes as $\Delta \sim N^{-1}$ at $\cpc=\pi/4$ (for $k=\pi$).
The winding number is $W=+1/2$ for $\lambda<\cpc$ and $W=-1/2$ for $\lambda>\cpc$, as
common to long-range pairing, $\alpha \leq 1$~\cite{PezzePRL2017, VodolaPRL2014}.
The symmetry under $\cp\to-\cp$ is lost, due to the loss of particle-hole symmetry.

\begin{figure}[t!]
\includegraphics[width=\columnwidth]{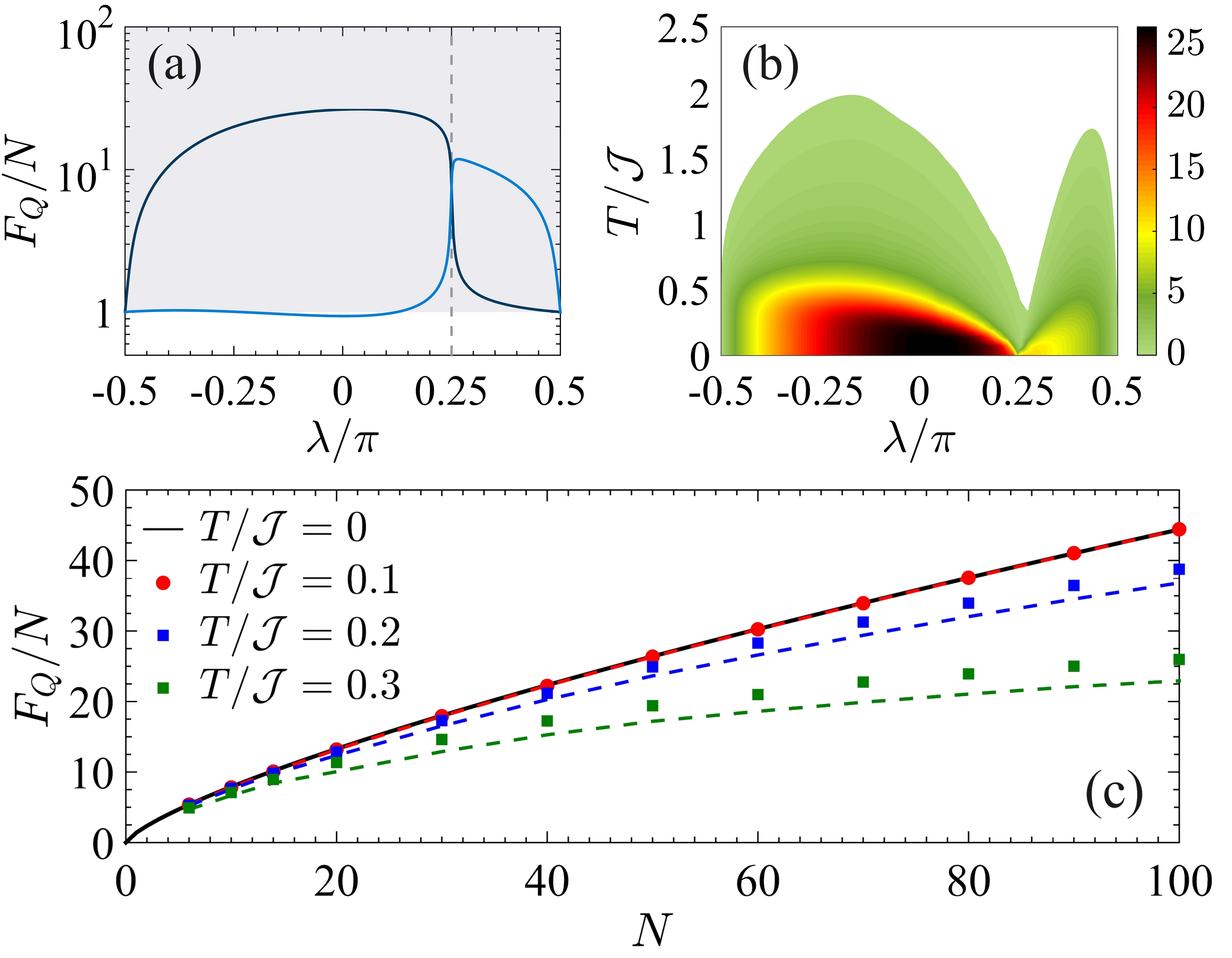}
\caption{
{\bf QFI for the Kitaev chain with long-range pairing.}
(a) Fisher density $F_Q[\ket{\psi_0}]/\sites$ as a function of $\cp$ for the ground state of Eq.~(\ref{HKitaev00}) with $N=50$ and $\alpha=0$.
The QFI is calculated using both the operators $\hat{O}_x^{+}$ (dark blue line)
and $\hat{O}_y^{(-)}$ (light blue line).
The vertical dashed line signals the critical point $\cpc$, while 
the shaded area marks multipartite entanglement. 
(b) Fisher density $F_Q[\hat{\rho}_T]/\sites$ (color scale) in the $\lambda$-$T$ plane for $N=50$.
(c) Scaling of $F_Q[\hat{\rho}_T]/\sites$ for increasing $N$.
The dashed lines are Eq.~(\ref{KITSRFish}) for different values of $T$.
} 
\label{fig10}
\end{figure}

In Figs.~\ref{fig9} and~\ref{fig10} we plot the QFI phase diagram and the witnessed multipartite entanglement, respectively. 
The operator that maximizes the QFI of the ground state is found to be $\hat{O}_x^{(+)}$ for $\cp \leq \cpc$ and $\hat{O}_y^{(-)}$ for $\cp \geq \cpc$, 
see Fig.~\ref{fig10}(a).
The two topological phases at $\cp<\cpc$ and $\cp>\cpc$ are characterized by a diverging multipartiteness $F_Q/\sites \sim \sites^{3/4}$, 
while $F_Q/\sites \sim \sites^{1/2}$ at criticality $\cp=\cpc$~\cite{PezzePRL2017}.
These scaling behaviors survive at low temperature as shown in Fig.~\ref{fig9}(a) where we plot 
$F_Q[\hat{\rho}_T]/F_Q[\ket{\psi_0}]$ in the $\lambda$-$T$ phase diagram.
For low temperatures $T\lesssim\Tcross$, we recognize two QPs at both sides of the critical point.
Since the transition is characterized by $\Delta_Q=1/2$ and $z=1$, the thermal decay 
$F_Q[\hat{\rho}_T] \sim T^{-1/2}$ for $T \gg \Delta$ characterizes the CP around $\cpc$.
In Fig.~\ref{fig9}(b), the green region highlights values $\beta \in [-0.53,-0.47]$.
Figure \ref{fig10}(b) highlights the region of the $\lambda$-$T$ phase diagram where the QFI witnesses multipartite entanglement (colored region).
In particular, in Fig.~\ref{fig10}(c) we plot $F_Q[\hat{\rho}_T]/N$ as a function of $N$ for $\lambda=0$ and different temperatures.
For sufficiently small temperature the QFI is bounded by Eq.~(\ref{QFIfactorization}). 
In this case, the evaluation of the degeneracy of the first excited state $\nu$ is not easily practicable:
an analysis of Eq.~(\ref{spect}) shows that the number of states in a small interval centered around the energy of the first excited state increases with $N$.
We thus superpose in Fig.~\ref{fig10}(c) the numerical data (dots) for $F_Q[\hat{\rho}_T]/N$ to the curve
$\tanh^2 \left(\frac{\Delta}{2T}\right) \frac{1+\neper^{-\Delta/T}}{1+cN\neper^{-\Delta/T}}$ (dashed lines) as suggested by Eq.~(\ref{QFIfactorization}), 
where $c=1.4$ is a fitting parameter and $\Delta = 0.91\Eunit$ (in the thermodynamic limit).
For $N \ll \neper^{\Delta/T}/c$, the QFI grows as $F_Q[\hat{\rho}_T]/N \sim N^{3/4}$.

\section{Conclusions}
\label{outlook}

The QFI, as a multipartite entanglement witness, allows to study strongly-correlated systems from a 
quantum information perspective and is thus attracting increasing interest~\cite{Strobel2014, HaukeNATPHYS2016, PezzePNAS2016, 
RajabpourPRD2017, LiuJPA2013, HaukeNATPHYS2016, PezzePRL2017, ZhangPRL2018, PappalardiJSM2017, FrerotPRB2016}. 
Differently from bipartite/pairwise entanglement measures
the QFI \cite{HaukeNATPHYS2016} (or close lower bounds \cite{Strobel2014, LueckeSCIENCE2011, BohnetSCIENCE2016}) 
can be extracted experimentally in arbitrary large systems 
of atomic ensembles and solid-state platforms. 

In this manuscript we have discussed the universal behavior of the QFI for systems at thermal equilibrium close to a QPT. 
At low-temperature, the QFI is lower bounded by a simple function that only depends on the structure of the two low-lying energy levels
and is factorable in a finite-temperature and a zero-temperature contributions.
This feature allows to draw a V-shaped phase diagram for the QFI centered at the critical point, Fig.~\ref{fig1}, which is  
common to both symmetry-breaking and topological QPTs.
We showed the existence of a universal low-temperature region -- the CP -- 
where thermal decay of the QFI is ruled by few fundamental critical exponents.
This region fans out from the critical point and can be identified as a quantum critical regime
where quantum coherence has a behavior controlled by the transition and competes with thermal fluctuations.
The universal behavior is lost at surprisingly high temperatures.

Finally, the analysis has emphasized the robustness of multipartite entanglement at finite temperature. 
In particular, a superextensive QFI (with a scaling at the Heisenberg limit $F_Q \sim N^2$) 
survives up to high temperatures, $T \propto 1/\log N$ in topological systems with large finite size.
This is an important difference with respect to models showing symmetry-breaking QPTs. 
In the latter systems multipartite entanglement is generally found at finite temperature in the disordered phase and 
the superextensive QFI that characterizes the ground state of the ordered phase is exponentially fragile 
against temperature, being lost for $T \propto \neper^{-N}$.  \\

{\bf Acknowledgments.}
We thank T. Roscilde and I. Fr\'erot for helpful discussions and for sharing their recent draft~\cite{FrerotDRAFT} 
where the quantum variance, a quantity related to the quantum Fisher information, is studied 
at finite temperature around the critical point of many-body quantum models and used to characterize a quantum critical regime.
We also acknowledge discussions with R. Franzosi, L. Lepori and M. Gessner. 
This work has been supported by the QuantERA projects ``Q-Clocks'' and ``TAIOL''. \\

\newpage


\begin{thebibliography}{400}
\bibitem{NOTA03}
For low temperatures $T\ll\DeltaE$, the QFI can be expanded in powers of the small quantity $\neper^{-\DeltaE/T}$.
For $k=2$ we obtain $F_Q[\hat{\rho}_T] /F_Q[\ket{\psi_0}] = 1-4\,\neper^{-\Delta/T} + \pazocal{O}(\neper^{-\Delta/T})^2$.
For $k>2$, the result $F_Q[\hat{\rho}_T] /F_Q[\ket{\psi_0}] = 1-2\,\neper^{-\Delta/T}+\pazocal{O}(\neper^{-\Delta/T})^2$ 
is independent on $k$.
Only in the limit $T\to0$ the two-mode approximation agrees with this expansion.
\bibitem{NOTA04}
The Fisher matrix Eq.~(\ref{QFIoptimization}) reads 
$\mathbb{F}_Q[\hat{\rho}_T]/N=\mathrm{diag}\big(0, \sqrt{1-\cot\lambda}, 1/\sqrt{1-\cot\lambda}\big)\times\tanh(\Delta_1/2T)$.
Thus, the optimal operator is $\hat{S}_z$ for all $T$.
\bibitem{NOTA05}
A finite-size analysis extended to $N=10\div10^3$ reveals that the energy gap vanishes 
exponentially $\Delta_1 \propto \neper^{-N/n_\DeltaE}$ for $\cp<\cpc$, 
with $n_\DeltaE = a|1-\cot\lambda|^b$ ($a\approx1.5$, $b\approx-0.85$).
\end{thebibliography}
\end{document}